\documentclass[a4paper,11pt]{article}
\pdfoutput=1

\usepackage{jheppub1}
\usepackage[T1]{fontenc}
\usepackage{amssymb}
\usepackage{graphicx}
\usepackage{amsmath}
\usepackage{hyperref}
\usepackage{tensor}
\usepackage{epstopdf}
\usepackage{extarrows}
\usepackage{verbatim}
\usepackage{subfigure}

\newcommand{\td}{\text{d}}

\begin{document}
	
	\title{Upper bounds of holographic entanglement entropy growth rate for thermofield double states}
	
	\author{Ze Li and Run-Qiu Yang}
	\emailAdd{lize@tju.edu.cn}
	\emailAdd{aqiu@tju.edu.cn}
	\affiliation{Center for Joint Quantum Studies and Department of Physics, School of Science, Tianjin University, Yaguan Road 135, Jinnan District, 300350 Tianjin, P.~R.~China}
	
	\abstract{We studied the upper bounds of the holographic entanglement entropy growth rate for thermofield double (TFD) states. By comparing the cases of vacuum AdS and charged AdS black holes,  we conjecture: for all static planar or spherically symmetric asymptotically Schwarzschild-AdS black holes of same mass density or entropy density, the vacuum AdS black hole gives the maximum entanglement entropy growth rate. We gave proofs by assuming dominant energy condition. We also considered the AdS black hole spacetime with real scalar fields case, where the scalar fields violate the dominant energy condition and the bulk geometry is not asymptotically Schwarzschild-AdS. Numerical results show that this case vacuum black hole still has maximal growth rate if we fixed entropy. However, in the case of fixed energy, vacuum case has maximal growth rate of entanglement entropy only under standard quantization scheme.
	}
	
	%
	%
	%
	\maketitle
	%
	%
	
\section{Introduction}
In recent years, exciting works based on the correspondence between the gravity of asymptotically anti-de Sitter spacetime and the conformal field theory, i.e. AdS/CFT correspondence, emerge one after another.
Roughly speaking, the AdS/CFT correspondence says that there is an equivalent relationship between $d+1$-dimensional gravity theory in asymptotically AdS$_{d+1}$ spacetime and $d$-dimensional conformal field theory on the boundary of the AdS$_{d+1}$ spacetime.
Since it was first proposed in \cite{Maldacena:1997re}, there have been more and more evidences to support it in various different models. In these models people have concluded a few of universal dual relationship between the quantities in gravity theory and boundary field theory.
	
The entanglement entropy is an important quantity in the studies of both gravity itself and the dual boundary field theory. In  quantum field theory and  quantum mechanical systems it is usually computed by the von Neumann entropy formula
\begin{equation}
	S_{vN}=-\rm Tr \rho \rm ln \rho,
\end{equation}
where $\rho$ is density matrix of a quantum system. In some situations, the von Neumann entropy is not easy to be calculated directly. However, we can calculate it by the AdS/CFT correspondence.
The Ryu-Takayanagi (RT) prescription tells us that the entanglement entropy of boundary CFT can be calculated by the R-T surface $\gamma_{A}$ in the bulk as~\cite{Ryu:2006bv}
\begin{equation}\label{RTF1}
	S_{A}=\frac{\rm{Area}(\gamma_{A})}{4G_{N}^{(d+1)}},
\end{equation}
here $\gamma_{A}$ is the codimension-2 bulk extremal surface homologous to subsystem $A$ of dual CFT\footnote{In the event that there are various extremal surfaces, the one with minimal area among them is picked. }. The $G_{N}^{(d+1)}$ is the $(d+1)$-dimensional Newton's constant. The original proposal~\eqref{RTF1} was given on a particular spacelike slice and is not covariant. As for covariant entanglement entropy, we have the Hubeny-Rangamani-Takayanagi (HRT) surface  correspondingly~\cite{Hubeny:2007xt}. Furthermore, if we take the contribution of bulk matters into account, it will require the extremal surface to minimize the generalized entropy
\begin{equation}
	S_{\rm{EE}}=\rm{min} \left\{\text{ext} \left[\frac{\text{Area}(\gamma_{A})}{4G_{N}^{(d+1)}}+S_{\rm bulk}(\Sigma_{\gamma_{A}})\right]\right\},
\end{equation}
where $\Sigma_{\gamma_{A}}$ is the region surrounded by $\gamma_{A}$ and boundary subsystem $A$.
The $\gamma_{A}$ that makes $S_{EE}$ to the minimum value is called the quantum extremal surface (QES)~\cite{Faulkner:2013ana,Engelhardt:2014gca}.
	
The AdS black hole spacetime of maximum analytical continuation is dual to a CFT with two copies on its two boundaries respectively~\cite{Israel:1976ur,Maldacena:2001kr}. The two equal-time boundary slices on the two boundary  are entangled with each other and together give us a dual description of a thermofield double(TFD) state. The reduced density matrix of time slice in every boundary  describes a thermal state. Usually, the extremal surface will not penetrate the horizon~\cite{Ryu:2006bv,Hubeny:2012ry}.
However, for the AdS black hole spacetime of maximum analytic continuation, the extremal surface has possibility to penetrate the horizon if its two "endpoints" are anchored on the two side boundaries respectively ~\cite{Hubeny:2007xt,Hartman:2013qma,Abajo-Arrastia:2010ajo,Aparicio:2011zy,Albash:2010mv,Balasubramanian:2010ce,Balasubramanian:2011ur}. Such surfaces are also called  Hartman-Maldacena(H-M) surfaces.
The H-M surface is an important tool for computing the holographic entanglement entropy. It has some special properties, such like it increases over time and the growth rate will approach a constant
when time goes to infinity~\cite{Hartman:2013qma}.

It is physical interesting to study the possible maximal growth rate of entanglement entropy in many situations. For example, in quantum computation theory, the quantum entanglement is regarded as a computing resource, which provides the computational speed-up~\cite{Ekert:1997wh,Jozsa:1997hc}. Studying the growth rate of entanglement entropy can help us understand how to reach the maximum entanglement state quickly.
In addition, entanglement entropy plays an important role in the black hole information paradox.
To interpret the Page Curve is recognized as an important task of solving black hole information paradox~\cite{Page:1993wv,Page:2013dx}.
Some recent studies show that we can explain the Page Curve from the perspective of holography, which says that a region called "island" will appear in the later period of black hole evaporation~\cite{Almheiri:2020cfm,Almheiri:2019qdq,Penington:2019kki,Chen:2020hmv}. 
In principle, the dynamics of degrees of freedom of gravity and matter fields should be considered together for the appearance of an island. We may first simplify the discussion by assuming that the dynamical gravity region in the boundary is negligible. Then the time of appearance of island can be found by our static bulk geometry. By studying the maximum growth rate of the entropy computed from H-M surface, we expect that such bounds could help us to estimate when the island transition occurs.
	
The evolution of holographic entanglement entropy has been studied in many works~\cite{Hubeny:2007xt, Rangamani:2016dms,Hartman:2013qma, Abajo-Arrastia:2010ajo, Dong:2016hjy, Liu:2013qca, Allais:2011ys, Albash:2010mv, Ziogas:2015aja}, and we are going to ask whether there is a universal upper bound for its growth rate. Similar to the  holographic entanglement entropy which is calculated by the area of codimension-2 extremal surface, the holographic complexity in ``complexity-volume'' (CV)-conjecture is calculated by the volume of codimension-1 extremal surface. The quantum computation theory gives an upper bound of the growth rate of complexity when the total energy of the system is fixed, which is known as the Lloyd's bound~\cite{Lloyd2000-jm}. There is also a bound proved to be applicable to holographic complexity in CV-conjecture~\cite{Yang:2019alh}. This inspires us that there may also be an upper bound for the growth rate of entanglement entropy if the energy of the system is fixed.

In order to find the maximum growth rate of entanglement entropy for TFD states in holography, we consider the entanglement between two half planes which locate at two AdS boundaries symmetrically. This paper first calculates the growth rate of holographic entanglement entropy for the cases of vacuum spherically/planar/hyperbolically symmetric black hole spacetime and corresponding charged black hole spacetime with fixed mass, entropy density and temperature. We find the that the entanglement growth rate of vacuum black holes are always faster than the corresponding charged black hole with fixed mass density and entropy density. Then we will give a conjecture: for static asymptotically planar/spherically symmetric Schwarzschild-AdS black holes, the vacuum black hole gives the upper bound of the growth rate under the conditions of fixed mass density and entropy density. Then we will give proofs under dominant energy condition. Furthermore, we perform numerical calculation for the case that the spacetime with real scalar fields. Here the dominant energy condition is not satisfied  and the bulk spacetime is asymptotically AdS but not asymptotically Schwarzschild-AdS. We find that the previous conclusion is still valid under one  of the two quantization schemes~\cite{Marolf:2006nd}.
	
This paper is organized as follows: We will introduce basics of TFD states and H-M surfaces in section \ref{section 2}.
Then we study the growth rate of holographic entanglement entropy in Schwarzschild-AdS black hole spacetime case and RN-AdS black hole spacetime cases as examples in section \ref{section 3}. We next give a general proof for the static spacetime with planar and spherical symmetries in section \ref{section 4} and assuming dominant energy condition. In section \ref{section 5}, we will consider spacetime with matter fields that do not obey the dominant energy condition and the bulk is asymptotically AdS but not Schwarzschild-AdS.

\section{Area of extremal surface and its growth rate}\label{section 2}
In this section, we first introduce our holographic model of black hole spacetime dual to boundary thermal CFTs. Then we will calculate the area of the extremal surface. We will also give a method to compute the growth rate of the area.
\subsection{The holographic model}
We consider the $(d+1)$-dimensional static AdS black hole spacetime with spherical/planar/hyperbolic symmetries as the bulk gravity theories, of which metric reads
\begin{equation}
	\label{eq:4051}
	\td s^2=\frac{1}{z^2}\left[-f(z) \mathrm{e} ^{-\chi(z)}\td t^2+\frac{\td z^2}{f(z)}+\td \Sigma_{k,d-1}^2\right].
\end{equation}
The $\td \Sigma_{k, d-1}^{2}$ is $(d-1)$-dimensional angular direction line element, which is given by
\begin{equation}
	\label{eq:4052}	
	\td \Sigma_{k, d-1}^{2}= \begin{cases}\td \Omega_{d-1}^{2}=\td \theta^{2}+\sin ^{2} \theta \td \Omega_{d-2}^{2} & \text { for } k=+1 \\  \td \mathbf{x}_{d-1}^2=\sum_{i=1}^{d-1} \td x_{i}^{2} / \ell_{\rm AdS}^{2} & \text { for } k=0 \\ \td \Xi_{d-1}^{2}=\td \sigma^{2}+\sinh ^{2} \sigma \td \Omega_{d-2}^{2} & \text { for } k=-1\end{cases}.
\end{equation}
Where $k=\{1,0,-1\}$ represent spherical, planar and hyperbolic symmetries of $(d-1)$-dimensional spatial directions respectively. The spatial direction coordinates $\theta \in [0,\pi]$ and $\sigma \in (-\infty,+\infty)$. The $\td\Omega_{d-2}^2$ is $(d-2)$-dimensional spherical coordinates metric. The $\ell_{\mathrm{AdS}}$ is the AdS radius. The maximum analytical continuation of the black hole spacetime described by metric (\ref{eq:4051}) is a two-sided black hole spacetime. It is dual to two copies of finite temperature CFTs defined on the left and right side boundaries. The temperature of the CFTs is just equal to the black hole temperature.
	
\begin{figure}
	\centering
	\includegraphics[width=0.75\textwidth]{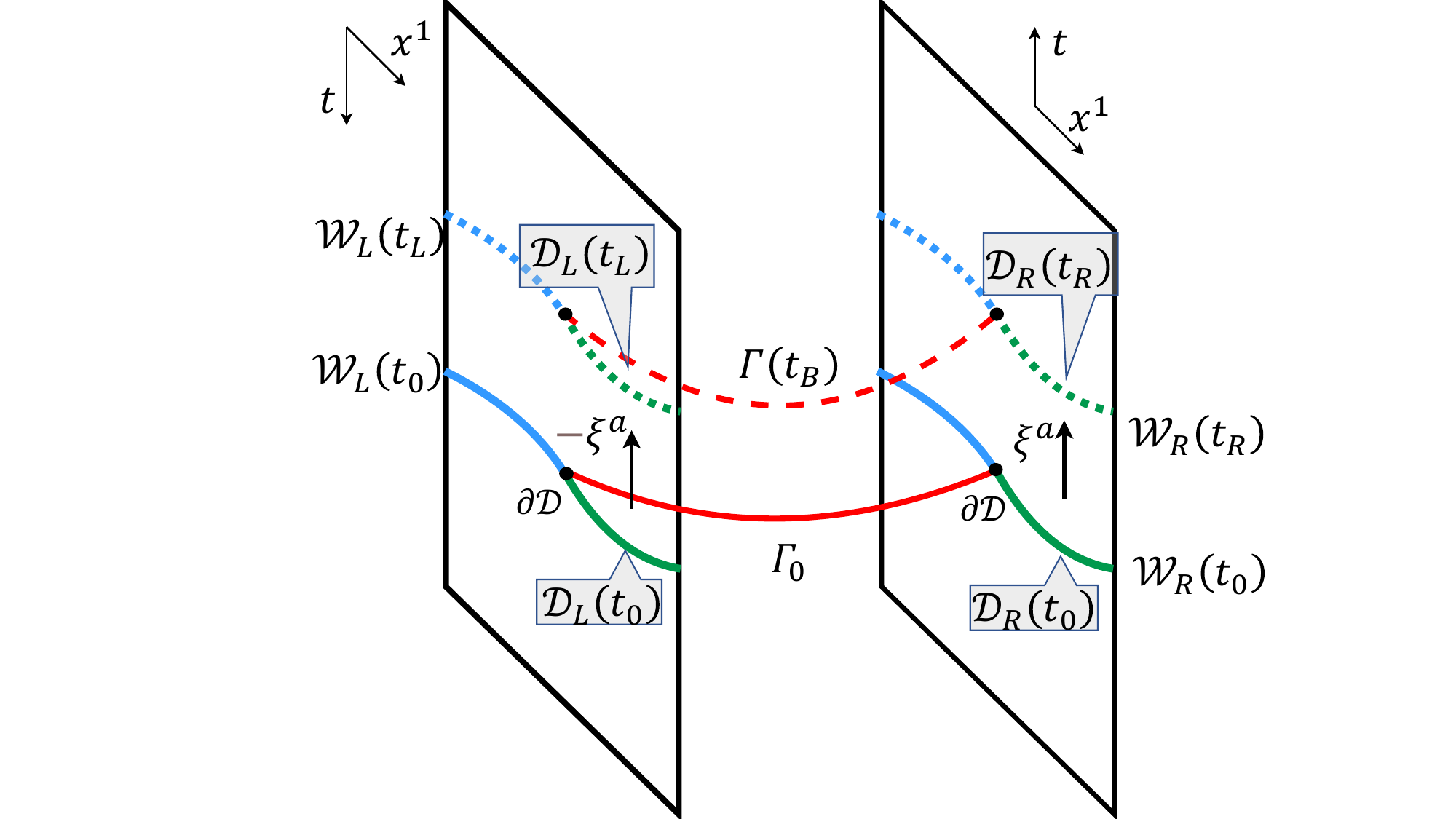}
	\caption{The sketch of boundary CFT time slice and extremal surface. The two boundary slices $\mathcal{W}_{R}$ and $\mathcal{W}_{L}$ combine into the total system $\mathcal{W}$. $\mathcal{D}$ is the subsystem of $\mathcal{W}$, which marked in green. The blue parts are the complement subsystem of $\mathcal{D}$. The extremal surface $\Gamma_{0}$ anchors on the subsystem boundaries $\partial \mathcal{D}$, which marked in red. $\xi^{a}$ and $-\xi^{a}$ are Killing vectors on the right and left boundaries.}\label{Fignull1}
\end{figure}

We first take a time slice $\mathcal{W}(t_{B})=\mathcal{W}_{R}(t_{B})\cup \mathcal{W}_{L}(t_{B})$ of the two-sided CFT as the total system. Here $t_{B}$ is the time coordinate on the boundaries.
Then we choose the subsystem  $\mathcal{D}(t_{B})=\mathcal{D}_{R}(t_{B})\cup \mathcal{D}_{L}(t_{B})$, where $\mathcal{D}_{R}(t_{B})$ and $\mathcal{D}_{L}(t_{B})$ are subsystems of the right and left side CFT time slice $\mathcal{W}_{R}(t_{B})$ and $ \mathcal{W}_{L}(t_{B})$ respectively. We set $\mathcal{D}_{R}(t_{B})$ and  $\mathcal{D}_{L}(t_{B})$ to satisfy
\begin{equation}
	\label{eq:40230}
	\begin{aligned}
		&\begin{cases}\mathcal{D}_{R}(t_{B})&=\{x^{1}\textgreater 0|\mathcal{W}_{R}(t_{B})\} \\  \mathcal{D}_{L}(t_{B})&=\{x^{1}\textgreater 0|\mathcal{W}_{L}(t_{B})\} \end{cases} \qquad \text { for } k=0, \\
		&\begin{cases}\mathcal{D}_{R}(t_{B})&=\{0 \textless \theta \textless \pi/2|\mathcal{W}_{R}(t_{B})\} \\  \mathcal{D}_{L}(t_{B})&=\{0 \textless \theta \textless \pi/2|\mathcal{W}_{L}(t_{B})\} \end{cases} \qquad \text { for } k=+ 1, \\
		&\begin{cases}\mathcal{D}_{R}(t_{B})&=\{0 <\sigma< +\infty|\mathcal{W}_{R}(t_{B})\} \\  \mathcal{D}_{L}(t_{B})&=\{0 < \sigma< +\infty|\mathcal{W}_{L}(t_{B})\} \end{cases} \qquad \text { for } k=- 1.
		\end{aligned}
\end{equation}
That is to say, we separate the $(d-1)-$dimensional space $\mathcal{W}$ into two half spaces.  The time evolution of the boundary subsystem $\mathcal{D}(t_{B})$ is non-trivial, since it is generated by two different time-like Killing vectors $\xi^{a}$ and $-\xi^{a}$ on the right and left boundaries respectively(see Figure \ref{Fignull1}), where $\xi^a$ is the bulk Killing vector which stands for the static symmetry. The entanglement entropy will also change over time, we can compute it by computing the area of extremal surface that anchored on the two boundaries $ \partial \mathcal{D}_{L}$ and $ \partial \mathcal{D}_{R}$ of subsystem $\mathcal{D}$. Such surface is known as the  Hartman-Maldacena(H-M) surface~\cite{Hartman:2013qma}.

In fact, for the case of spherical symmetry, the extremal surface may not pass through the event horizon which is disconnected. The entanglement entropy of the disconnected phase is related to the thermal entropy of the black hole~\cite{Hubeny:2013gta}. Which of the Hartman-Maldacena saddle and the thermal saddle gives the minimal extremal surface depends on the setting of parameters(horizon radius, AdS radius, size of subregion, etc). The extremal surface of thermal saddle does not evolve with time. Its time derivative is zero, which is always less than the positive growth rate of the Hartman-Maldacena saddle. When discussing the upper bound on the growth rate below, we only need to discuss the Hartman-Maldacena saddle, since the thermal saddle will naturally satisfy the upper bound of growth rate we mentioned later.

\subsection{Area of extremal surface and its Growth rate}
\begin{figure}
	\centering
	\includegraphics[width=0.65\textwidth]{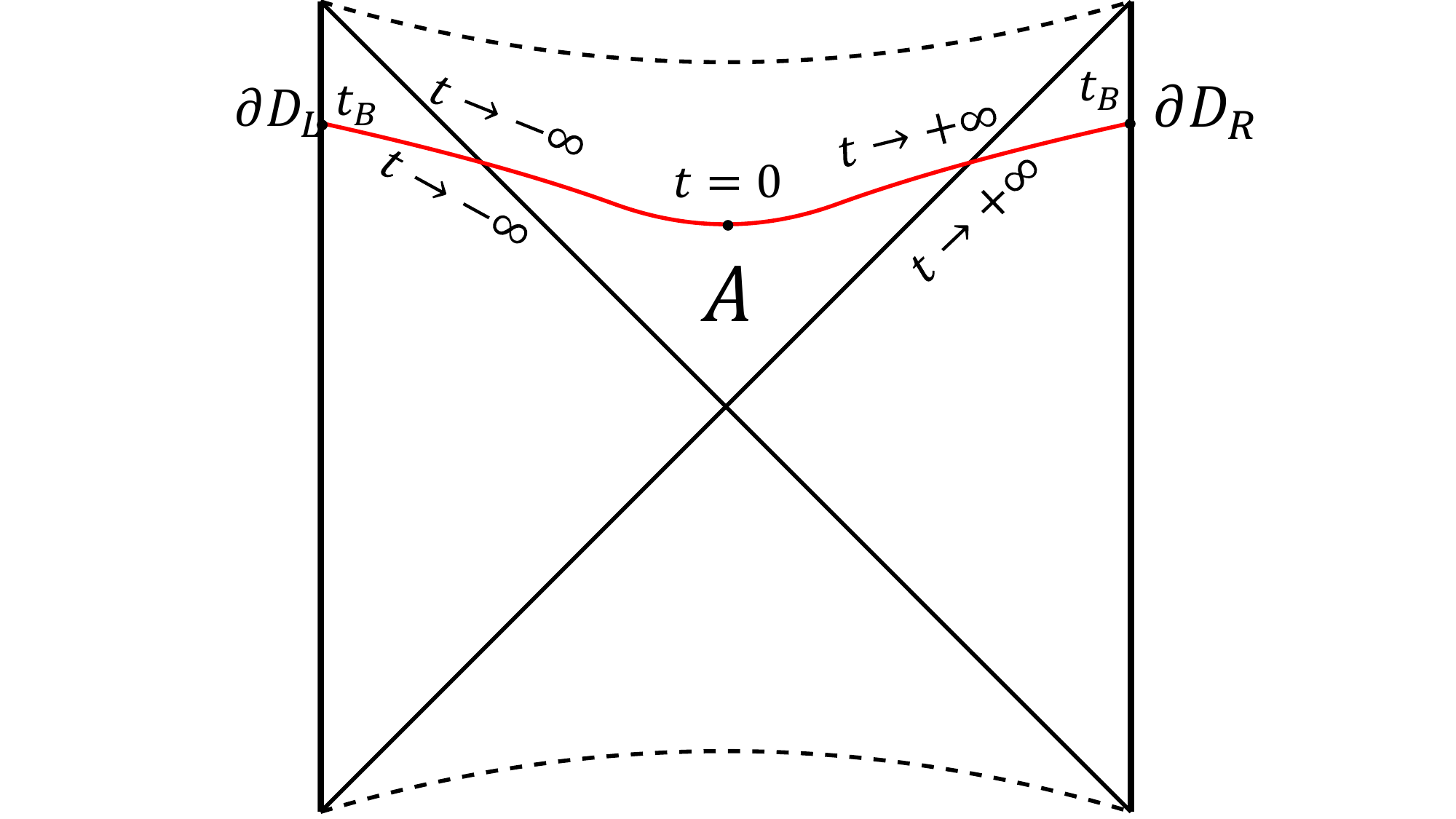}
	\caption{Set $t_{R}=-t_{L}=t_{B}$. A is the middle point with $t=0$. } \label{Fignull13}
\end{figure}

Since we can always let $t_{R}=-t_{L}=t_{B}\geqslant 0$ by Lorentz boost, the extremal surface could be symmetric in the left and right parts. Due to the symmetry, this extremal surface can be parameterized locally by
\begin{equation}\label{HTsurface}
	z=z(\lambda),t=t(\lambda)\,.
\end{equation}
with the boundary condition $z(-\infty)=z_{L}=0, z(\infty)=z_{R}=0, t(-\infty)=t_{L}=-t_B$ and $t(\infty)=t_{R}=t_B$. Also due to the symmetry, we can see that there is ``point'' (Strictly speaking, is codimensional-2 sub-manifold) where $z_A$ is local maximal and $t_A=0$. See Figure \ref{Fignull13}. This means that $\partial z/\partial t|_A=0$. Let us denote the area of extremal surface $\Gamma$ to be $A(\Gamma)$. Using the ``point'' A we can seperate the extremal surface $\Gamma$ into two symmetric parts. The area $\mathcal{A}(\Gamma)$ can be computed by the right half part multiplied by two. For convenience, we eliminate the parameter $\lambda$ and rewrite the $z$-coordinate as the function of $t$ locally. At the right part, since the $t$-coordinates first runs from zero to infinity and then runs from infinity to $t_B$, the function $z=z(t)$ should separated into two parts. The area of the extremal surface $\Gamma$ reads
\begin{equation}
	\label{eq:4054}
	\mathcal{A}(\Gamma)=2V_{k,d-2}\left(\int_{0}^{+\infty} \td t \frac{1}{z^{d-1}}\sqrt{-f(z)\mathrm{e} ^{-\chi(z)}+\frac{\dot{z}^2}{f(z)}}+\int_{+\infty}^{t_{B}} \td t \frac{1}{z^{d-1}}\sqrt{-f(z)\mathrm{e} ^{-\chi(z)}+\frac{\dot{z}^2}{f(z)}}\right).
\end{equation}
Where $V_{k,d-2}=\int\td \sigma_{k,d-2}$ is the unit volume of the $(d-2)-$dimensional spatial directions, here $\td \sigma_{k,d-2}^{2}$ is the $(d-2)-$dimensional line element produced by $\td \Sigma_{k,d-1}^{2}$ fixing one coordinate $x_{1}$ or $\theta$. The $V_{k,d-2}$ will be divergent with $k=0,-1$, but we can set it as a large constant by taking a cut off, this does not change our final results\footnote{We can also define a area density by dividing the area by $V_{k,d-2}$ to eliminate divergence.}. The integration of Eq.~\eqref{eq:4054} will be divergent since it goes to the bulk boundary, but we will see later that its derivative with respect to time is finite.

\begin{figure}
	\centering
	\includegraphics[width=0.5\textwidth]{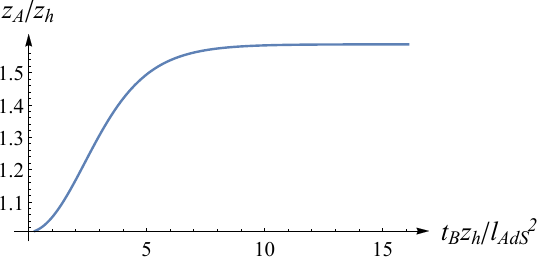}
	\includegraphics[width=0.45\textwidth]{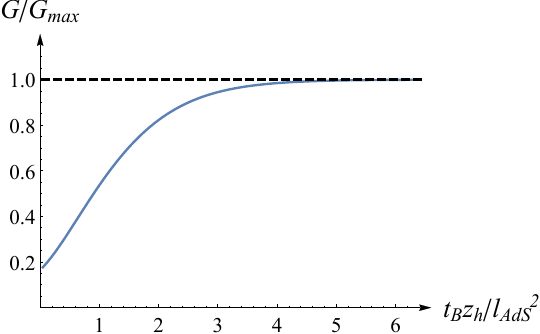}
	\caption{Left: The image of  $z_{A}$ and $t_{B}$. Right: The image of $G(z_{A})$ and $t_{B}$. Here we set $d=3$, $\ell_{\rm AdS}=100$, and $f(z)=h(z)$ as \ref{eq:40521} with $k=0$. } \label{Fignull131}
\end{figure}

\begin{figure}
	\centering
	\includegraphics[width=0.5\textwidth]{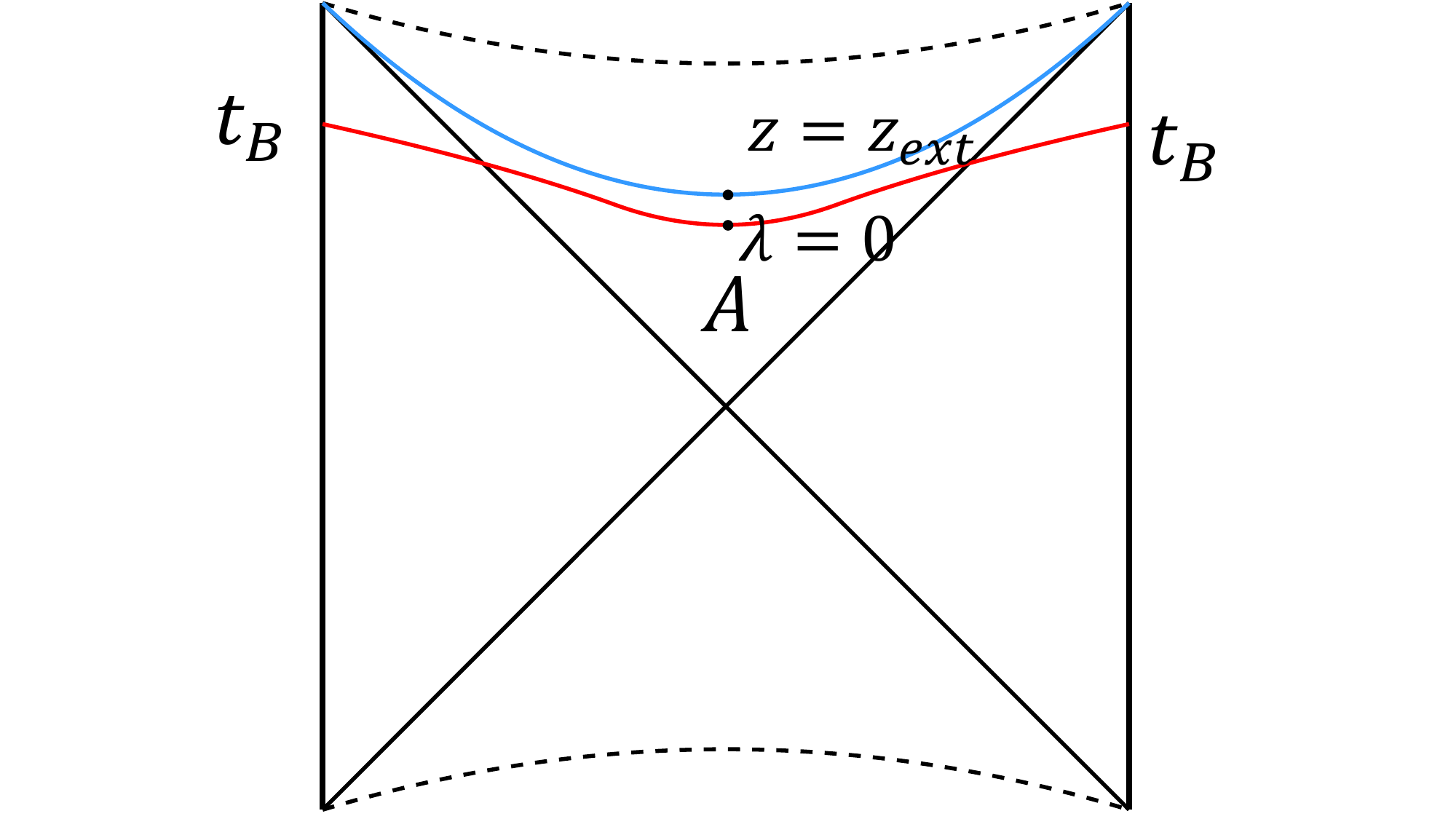}
	\caption{The red curve is the extremal surface, the blue curve is the extremal surface with $z_{A}=z_{\rm ext}$. With the time $t_{B}$ goes to infinite, the extremal surface will be infinitely approach to the blue surface.} \label{Fignull14}
\end{figure}

In order to get the growth rate of area, we can regard the area $\mathcal{A}$ as an action, and the integrand function of (\ref{eq:4054}) as a Lagrangian, which is a functional of generalized coordinate and generalized velocity $z,\dot{z}$
\begin{equation}
	\label{eq:40542}
	\mathcal{A}=2V_{k,d-2}\int\mathcal{L}(z,\dot{z})\mathrm{d}t,~~\mathcal{L} (z,\dot{z})=\frac{1}{z^{d-1}}\sqrt{-f(z)\mathrm{e} ^{-\chi(z)}+\frac{\dot{z}^2}{f(z)}}.
\end{equation}
According to the Hamilton-Jacobi equation of classical mechanics, the partial derivative of on-shell action $S$ over time $\partial S/\partial t$ is negative Hamiltonian, so the partial derivative of area over time is $\partial \mathcal{A}/\partial t=-2V_{k,d-2}\mathcal{H}$. Where $\mathcal{H}$ is the "Hamiltonian" of Lagrangian~\eqref{eq:40542}.
At the boundary we have $z=0$, so that $t=t_{B}$, we get the rate of area growth over the boundary time
\begin{equation}
	\label{eq:40501}
	\frac{\td \mathcal{A}}{\td t_{B}}=\left.-2V_{k,d-2}\mathcal{H}\right|_{t=t_B,z=0}.
\end{equation}
Since the Lagrangian $\mathcal{L}$ does not contain generalized coordinate $t$, we can get a conserved quantity $\mathcal{H}$ on the extremal surface
\begin{equation}
	\label{eq:4055}
	\mathcal{H}=\frac{\partial \mathcal{L}}{\partial \dot{z}}\dot{z}-\mathcal{L}=\frac{f(z)\mathrm{e} ^{-\chi(z)}}{z^{d-1}\sqrt{-f(z)\mathrm{e} ^{-\chi(z)}+\frac{\dot{z}^2}{f(z)}}}.
\end{equation}
This ``Hamiltonian'' is constant along the extreme surface. In order to figure out the conserved quantity $\mathcal{H}$, we focus on the middle point A with $z=z_{A}$, where we have $\dot{z}=0$. Then we can obtain
	
\begin{equation}
	\label{eq:4059}
	\mathcal{H}=-\frac{\sqrt{-f(z_{A})\mathrm{e} ^{-\chi(z_{A})}}}{z^{d-1}_{A}}.
\end{equation}
Since (\ref{eq:4055}) and (\ref{eq:4059}) are both conserved quantity on the extremal surface, we can get an equation
\begin{equation}
	\label{eq:40591}
	-\frac{\sqrt{-f(z_{A})\mathrm{e} ^{-\chi(z_{A})}}}{z^{d-1}_{A}}=\frac{f(z)\mathrm{e} ^{-\chi(z)}}{z^{d-1}\sqrt{-f(z)\mathrm{e} ^{-\chi(z)}+\frac{\dot{z}^2}{f(z)}}}.
\end{equation}
Solve this equation, we can get
\begin{equation}
	\label{eq:40592}
	\frac{\td t}{\td z}=\frac{\sqrt{-f(z_{A})\mathrm{e} ^{-\chi(z_{A})}}}{z^{d-1}_{A}}\frac{1}{f(z)\mathrm{e} ^{-\frac{\chi(z)}{2}}\sqrt{\frac{-f(z_{A})\mathrm{e} ^{-\chi(z_{A})}}{z^{2(d-1)}_{A}}+\frac{f(z)\mathrm{e} ^{-\chi(z)}}{z^{2(d-1)}}}}.
\end{equation}
For convenience, we define a function $G(z)$ (which is also very important in the following) as
\begin{equation}
	\label{eq:40593}
	G(z)\equiv\frac{\sqrt{-f(z)\mathrm{e} ^{-\chi(z)}}}{z^{d-1}}.
\end{equation}
Integrate (\ref{eq:40592}) over $z$, we get the boundary time $t_{B}$, which is a function of $z_{A}$
\begin{equation}
	\label{eq:40594}
	\begin{aligned}
		t_{B}(z_{A})&=\int_{z_{A}}^{0}\frac{G(z_{A})}{f(z)\mathrm{e} ^{-\frac{\chi(z)}{2}}\sqrt{G(z_{A})^2+\frac{f(z)}{z^{2(d-1)}}}} \td z.\\
		&=\int_{z_{A}}^{z_{h}+\epsilon}\frac{G(z_{A})}{f(z)\mathrm{e} ^{-\frac{\chi(z)}{2}}\sqrt{G(z_{A})^2+\frac{f(z)}{z^{2(d-1)}}}} \td z\\
		&-\int_{0}^{z_{h}-\epsilon}\frac{G(z_{A})}{f(z)\mathrm{e} ^{-\frac{\chi(z)}{2}}\sqrt{G(z_{A})^2+\frac{f(z)}{z^{2(d-1)}}}} \td z.
	\end{aligned}
\end{equation}
Here we introduce a cut off $\epsilon$, since the integral diverges at the horizon. The divergences at two sides just cancel with each other, so the integral obtains a finite contribution when it pass through horizon. Figure \ref{Fignull131} is obtained by numerical calculating $t_{B}$ by (\ref{eq:40594}), which shows that boundary time $t_{B}$ goes to infinite with $z_{A}$ limit to a finite value. At the same time, the growth rate increases monotonically to its maximum. In the case that $G(z)$ has only one extreme point, the integral \eqref{eq:40594} is finite as long as $z_{A} \neq z_{\rm ext}$, where $z_{\rm ext}$ is the extreme point of $G(z)$. And when $z_{A} = z_{\rm ext}$, $t_{B}\rightarrow \infty$. We give a proof of this in Appendix~\ref{appendix A0}.
This means the extremal surface will be infinitely close to the surface $z=z_{\rm ext}$ for very late times (see Figure \ref{Fignull14}). If $G(z)$ has multiple one extreme points, saying $\{z_{\text{ext},1}, z_{\text{ext},2},z_{\text{ext},3},\cdots\}$ with $z_{\text{ext},1}<z_{\text{ext},2}<z_{\text{ext},3}<\cdots$, $z$ will approach to the first extreme point $z_{\text{ext},1}$ but may not the $z_{\text{ext}}$ when $t_B\rightarrow\infty$.

Consider the conserved quantity $\mathcal{H}$ in (\ref{eq:4059}) and the function defined in (\ref{eq:40593}), the growth rate finally is written as
\begin{equation}
	\label{eq:405012}
	\frac{\td \mathcal{A}}{\td t_{B}}=2V_{k,d-2}G(z_{A}).
\end{equation}
We can also use a single parameter $\lambda$ to parameterize the extremal surface and obtain the a same result, which is shown in Appendix~\ref{appendix A}.
The growth rate can be determined by $z$ coordinate of the middle point $A$. The $z_{A}$ will be a constant at late time as we have discussed before. This means the area of the extremal surface will growth linearly.

To find maximal growth rate of entanglement entropy, one needs to find the extremum of $G(z)$(at this time $z=z_{\rm ext}$). The derivative of $G(z)$ reads
	
\begin{equation}
	\label{eq:405103}
	G^{\prime}(z)=(1-d) z^{-d} \sqrt{-\mathrm{e} ^{-\chi(z)}f(z)}+\frac{z^{1-d} \mathrm{e} ^{-\chi(z)}\left[f(z) \chi '(z)- f'(z)\right]}{2\sqrt{-\mathrm{e} ^{-\chi(z)}f(z)}}=0.
\end{equation}
then we get
\begin{equation}
	\label{eq:405104}
	2(d-1)f(z_{\rm ext})-z_{\rm ext}f^{\prime}(z_{\rm ext})+z_{\rm ext}f(z_{\rm ext}) \chi^{\prime}(z_{\rm ext})=0.
\end{equation}
For a given $f(z)$, we can figure out a corresponding $z_{\rm ext}$. It is clear that $z_{h}\textless z_{\rm ext}\textless +\infty$, i.e., the extreme point is inside the horizon but does not touch the singularity.

\section{Examples of vacuum black holes and charged black holes}\label{section 3}
We have given the growth rate of extreme surface area in the previous section, now we consider some examples to calculate and compare the growth rates. We will study the cases of vacuum  and charged black holes with different symmetries and fixed mass density, entropy density and temperature respectively.
\subsection{Schwarzschild black holes}
For Schwarzschild black hole, the metric is given by (\ref{eq:4051}). And the $f(z)$ and $\chi(z)$ is given by
\begin{equation}
	\label{eq:40521}
	f(z)=kz^{2}+\frac{1}{\ell_{\rm AdS}^{2}}-f_{0}z^{d},\qquad \chi(z)=0.
\end{equation}
Here $f_{0}$  is a parameters related to black hole mass which will be shown in the later formula (\ref{eq:405043}). For $k=0,1$, the $f_{0}$ is required to be positive, but for $k=-1$, $f_{0}$ could be negative, since we require the metrics are black hole solutions. Now we let $f(z_{h})=0$, which can give us  $f_{0}$ as
\begin{equation}
	\label{eq:405042}
	f_{0}=\frac{1}{z_{h}^{d-2}}\left( \frac{1}{z_{h}^{2}\ell_{\rm AdS}^{2}}+k\right),
\end{equation}
here $z_{h}$ is the $z$ coordinate of the horizon. From the metric, we can obtain the mass $M$, Bekenstein-Hawking entropy $S$, and temperature $T$ of the black hole
\begin{equation}
	\label{eq:405043}
	\begin{aligned}
		M&=\frac{(d-1)V_{k,d-1}}{16 \pi G_{N}}f_{0}=\frac{(d-1)V_{k,d-1}}{16 \pi G_{N}}\frac{1}{z_{h}^{d-2}}\left( \frac{1}{z_{h}^{2}\ell_{\rm AdS}^{2}}+k\right),\\
		S&=\frac{V_{k,d-1}}{4 G_{N}} z_{h}^{1-d},\\
		T&=-\frac{1}{4\pi}  \frac{\partial f(z)}{\partial z} \bigg|_{z=z_{h}}=\frac{1}{4\pi}\left((d-2)kz_{h}+\frac{d}{z_{h}\ell_{\rm AdS}^{2}}\right).
	\end{aligned}
\end{equation}
Where $V_{k,d-1}=\int\td \Sigma_{k,d-1}$ is the unit volume of the $d-1$ dimensional spatial directions.
	
The function $G(z)$ defined in (\ref{eq:40593}) will be $G(z)=\sqrt{-f(z)}/z^{d-1}$, and the maximum of $G (z)$ satisfies
\begin{equation}
	\label{eq:405044}
	2(d-1)f(z_{\rm ext})-z_{\rm ext}f^{\prime}(z_{\rm ext})=0.
\end{equation}
Substitute $f(z)$ into (\ref{eq:405044}), we obtain
\begin{equation}
	\label{eq:405045}
	2kz_{\rm ext}^{2}+\frac{2(d-1)}{(d-2)}\frac{1}{\ell_{\rm AdS}^{2}}-f_{0}z_{\rm ext}^{d}=0.
\end{equation}
So the maximum growth rate $G(z_{\rm ext})$ can be written as
\begin{equation}
	\label{eq:405046}
	G(z_{\rm ext})=\frac{\sqrt{kz_{\rm ext}^{2}+\frac{d}{d-2}\frac{1}{\ell_{\rm AdS}^{2}}}}{z_{\rm ext}^{d-1}}.
\end{equation}
The mass, Bekenstein-Hawking entropy, and temperature are given according to Eq.~\eqref{eq:405043}, we can fix each of them respectively to study the maximal growth rate with different topologies.
	
\subsubsection{Fixed mass density}
We will calculate and compare the growth rate of the cases with different $k$ with fixed mass density in this subsection. The first line of (\ref{eq:405043}) shows that, the mass $M$ is divergent in the cases of $k=0$ and $k=-1$. For different $k$, only the volume factor $V_{k,d-1}$ is different. We define the "mass density" $\mathcal{M}=M/V_{k,d-1}$. We keep $\mathcal{M}$ fixed, i.e., $f_{0}$ is fixed, and then compare the growth rates of the cases with different $k$.
	
When $k=0$, we have $f(z)=\frac{1}{\ell_{\rm AdS}^{2}}-f_{0}z^{d}$, then we can get the solution of (\ref{eq:405045})
\begin{equation}
	\label{eq:40505}
	z_{\rm ext}=z_{p\rm ext}:=\left[\frac{2(d-1)}{(d-2)\ell_{\rm AdS}^2f_{0}}\right]^{\frac{1}{d}}=\left[\frac{2(d-1)}{(d-2)}\right]^{\frac{1}{d}}z_{h}.
\end{equation}
For convenience, we define a coefficient $\gamma$ as $\gamma=2(d-1)/(d-2)$. Substitute (\ref{eq:40505}) into (\ref{eq:405045}), we can obtain
\begin{equation}
	\label{eq:40506}
	G_{p}(z_{\rm ext})=\sqrt{\frac{d}{d-2}} \gamma^{\frac{1}{d}-1} \ell_{\rm AdS}^{1-\frac{2}{d}} f_{0}^{1-\frac{1}{d}}.
\end{equation}
Here we use a footmark "$p$" to mark the growth rate of the vacuum planar symmetry case. The growth rate can be written as
\begin{equation}
	\label{eq:40507}
	\frac{\td \mathcal{A}}{\td t_{B}}=2V_{k,d-2}G_{p}(z_{\rm ext})=2V_{k,d-2}\sqrt{\frac{d}{d-2}} \gamma^{\frac{1}{d}-1} \ell_{\rm AdS}^{1-\frac{2}{d}} f_{0}^{1-\frac{1}{d}},
\end{equation}
which only depends on $V_{k,d-2}$, $d$, $\ell_{\mathrm{AdS}}$, and $f_{0}$.

For the spherically or hyperbolically symmetric Schwarzschild-AdS black hole, the function $f(z)$ is $f(z)=kz^{2}+\frac{1}{\ell_{\rm AdS}^{2}}-f_{0}z^{d}$, with $k=1$ or $k=-1$. Then we find the solution of (\ref{eq:405045}) satisfies
\begin{equation}
	\label{eq:40508}
	z_{\rm ext}^{d}=\frac{2kz_{\rm ext}^{2}}{f_{0}}+\frac{2(d-1)}{(d-2)\ell_{\rm AdS}^2f_{0}}.
\end{equation}
Noting Eq.~ \eqref{eq:40505}, we find that the (\ref{eq:40508}) can be written as
\begin{equation}
	\label{eq:405081}
	z_{\rm ext}^{d}=\frac{2kz_{\rm ext}^{2}}{f_{0}}+z_{p\rm ext}^{d}.
\end{equation}
Substituting (\ref{eq:405081}) into the expression of $G(z_{\rm ext})$, we can eliminate the $k$ and obtain the growth rate of the spherically or hyperbolically symmetric case
\begin{equation}
	\label{eq:405083}
	G(z_{\rm ext})=\frac{\sqrt{-f(z_{\rm ext})}}{z_{\rm ext}^{d-1}}=\frac{\sqrt{\frac{f_{0}}{2}(z_{\rm ext}^{d}+z_{p\rm ext}^{d})-\frac{1}{\ell_{\rm AdS}^2}}}{z_{\rm ext}^{d-1}}.
\end{equation}
	
We first consider the case of $k=1$. At this case, we have $z_{\rm ext} \textgreater z_{p\rm ext}$, then we get
\begin{equation}
	\label{eq:405084}
	G(z_{\rm ext})\textless \frac{\sqrt{f_{0}z_{\rm ext}^{d}-\frac{1}{\ell_{\rm AdS}^2}}}{z_{\rm ext}^{d-1}}.
\end{equation}
For the planar case, we have already known that $G_{p}(z_{p\rm ext})= \sqrt{f_{0}z_{p\rm ext}^{d}-1/\ell_{\rm AdS}^2}/z_{p\rm ext}^{d-1}$.
It's easy for us to know that function $G(x)=\sqrt{f_{0}x^{d}-1/\ell_{\rm AdS}^2}/x^{d-1}$ with $f_{0}x^{d}-1/\ell_{\rm AdS}^2\textgreater 0$(Corresponding to the $f(z)$ inside the horizon) and $d\geqslant 3$ is a monotone decreasing function. We compare $G(z_{\rm ext})$ and $G_{p}(z_{p\rm ext})$, combine the fact $z_{\rm ext} \textgreater z_{p\rm ext}$, and finally obtain
\begin{equation}
	\label{eq:405086}
	G(z_{\rm ext})\textless \frac{\sqrt{f_{0}z_{\rm ext}^{d}-\frac{1}{\ell_{\rm AdS}^2}}}{z_{\rm ext}^{d-1}}\textless  \frac{\sqrt{f_{0}z_{p\rm ext}^{d}-\frac{1}{\ell_{\rm AdS}^2}}}{z_{p\rm ext}^{d-1}}=G_{p}(z_{p\rm ext}).
\end{equation}
That is to say, the growth rate of spherical symmetry case is less than planar symmetry case with the same $f_{0}$.
	
\begin{figure}
	\centering
	\subfigure[$d=3$]{
		\includegraphics[width=0.45\textwidth]{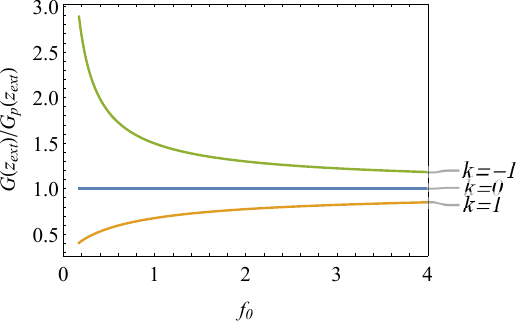}}\qquad
	\subfigure[$d=4$]{
		\includegraphics[width=0.45\textwidth]{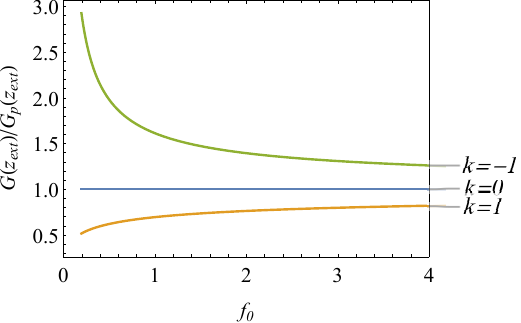}}

	\subfigure[$d=5$]{
		\includegraphics[width=0.45\textwidth]{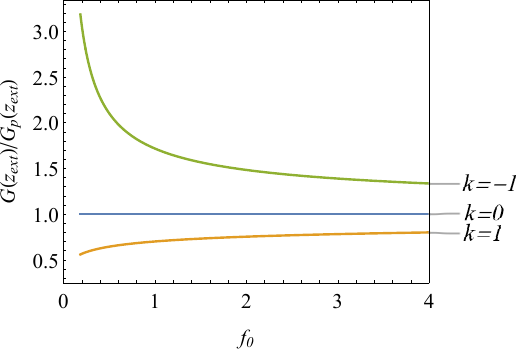}}\qquad
	\subfigure[$d=6$]{
		\includegraphics[width=0.45\textwidth]{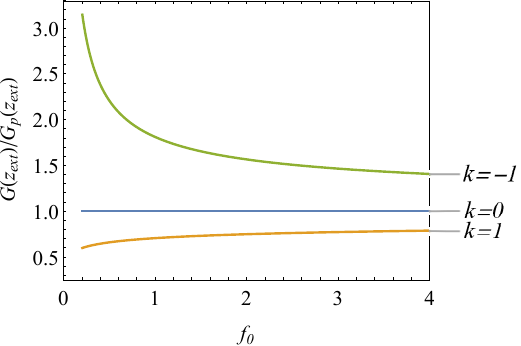}}
	\caption{For $\ell_{\rm AdS}=1$, the vertical axis represents $G(z_{\rm ext})/G_p(z_{\rm ext})$ and the horizontal axis represents $f_{0}$.}\label{Fignull7}
\end{figure}

For the $k=-1$ case, we find that we cannot analytically calculate and compare the $G(z_{\rm ext})$ with other case.  We numerically calculated the growth rate and obtained Figure \ref{Fignull7}.
From Figure \ref{Fignull7} we can see, with the same $f_{0}$, $G(z_{\rm ext})$ with $k=-1$ is larger than $G(z_{\rm ext})$ with $k=0$, and $G(z_{\rm ext})$ with $k=0$ larger than $G(z_{\rm ext})$ with $k=1$.
	
\subsubsection{Fixed entropy density}
We will calculate and compare the growth rate of different $k$ with fixed entropy density in this subsection. The second line of (\ref{eq:405043}) shows that, the entropy $S$ is also divergent in the cases of $k=0$ and $k=-1$. We can also define the "entropy density" $\mathcal{S}=S/V_{k,d-1}$.  The $\mathcal{S}$ only depends on $z_{h}^{1-d}$. So keeping the entropy density $\mathcal{S}$ means to keep $z_{h}$ constant. The function $f_{0}$ will be determined by (\ref{eq:405042}).
	
For the planar case($k=0$), the maximum of $G(z)$ is also given by (\ref{eq:40506})
	\begin{equation}
		\label{eq:4050431}
		G(z_{\rm ext})=\sqrt{\frac{d}{d-2}} \gamma^{\frac{1}{d}-1} \ell_{\rm AdS}^{-1} z_{h}^{1-d}=4G_{N} \gamma^{\frac{1}{d}-1}z_{1}\mathcal{S}
\end{equation}
where
\begin{equation}\label{defz1}
		z_{1}=\frac{1}{\ell_{\rm AdS}}\sqrt{\frac{d}{d-2}}.
\end{equation}

\begin{figure}
	\centering
	\subfigure[$d=3$]{
		\includegraphics[width=0.45\textwidth]{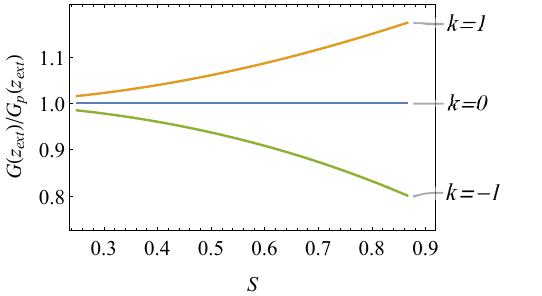}}\qquad
	\subfigure[$d=4$]{
		\includegraphics[width=0.45\textwidth]{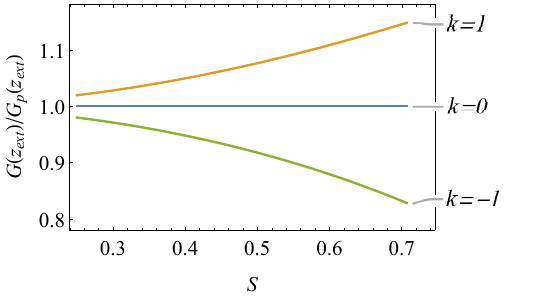}}

	\subfigure[$d=5$]{
		\includegraphics[width=0.45\textwidth]{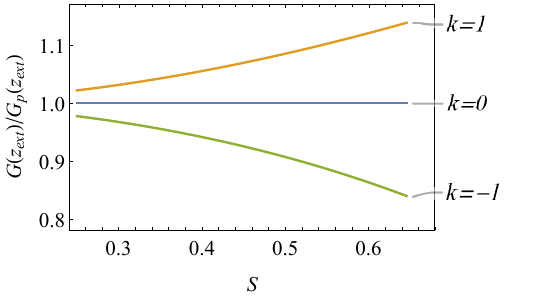}}\qquad
	\subfigure[$d=6$]{
		\includegraphics[width=0.45\textwidth]{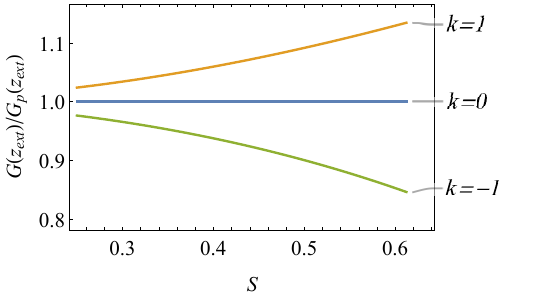}}
	\caption{For $\ell_{\rm AdS}=1$, the vertical axis represents $G(z_{\rm ext})/Gp(z_{\rm ext})$ and the horizontal axis represents $\mathcal{S}$.} \label{Fignull8}
\end{figure}

When $k\neq 0$,  there are no analytical expressions of $z_{\rm ext}$ and $G(z_{\rm ext})$. To compare $G(z_{\rm ext})$ with different $k$, we can treat $k$ as continuous real number, and then calculate the derivative of $G(z_{\rm ext})$ with respect to $k$ in the interval $k \in [-1,1]$
\begin{equation}
	\label{eq:4050432}
	\begin{aligned}
		\frac{\td G(z_{\rm ext})}{\td k}&=\frac{1}{2G(z_{\rm ext})}\frac{\td G^{2}(z_{\rm ext})}{\td k}\\
		&=\frac{1}{2G(z_{\rm ext})}\left[\frac{\partial G^{2}(z_{\rm ext})}{\partial k}+2G(z)\frac{\partial G(z)}{\partial z}\frac{\td z_{\rm ext}}{\td k}\Big|_{z=z_{\rm ext}}\right].
	\end{aligned}
\end{equation}
Since we have $\frac{\partial G(z)}{\partial z}\Big|_{z=z_{\rm ext}}=0$, (\ref{eq:4050432}) can be simplified to
\begin{equation}
	\label{eq:4050433}
	\frac{\td G(z_{\rm ext})}{\td k}=\frac{1}{2G(z_{\rm ext})}\frac{\partial G^{2}(z_{\rm ext})}{\partial k}=\frac{1}{2G(z_{\rm ext})}\left(\frac{\td f_{0}}{\td k}z_{\rm ext}^{2-d}-z_{\rm ext}^{4-2d}\right).
\end{equation}
Since we have $f(z_{h})=kz_{h}^{2}+\frac{1}{\ell_{\rm AdS}^{2}}-f_{0}z_{h}^{d}=0$, and $z_{h}$ is a constant. So we can obtain $\frac{\td f_{0}}{\td k}=z_{h}^{2-d}$. So the (\ref{eq:4050433}) becomes to
\begin{equation}
	\label{eq:4050434}
	\frac{\td G(z_{\rm ext})}{\td k}=\frac{z_{\rm ext}^{2-d}}{2G(z_{\rm ext})}\left(z_{h}^{2-d}-z_{\rm ext}^{2-d}\right).
\end{equation}
Since we always have $z_{h}\textless z_{\rm ext}$ and $d\textgreater 2$, the RHS of (\ref{eq:4050434}) is obviously greater than zero, which means $G(z_{\rm ext})$ increases as $k$ increases. That is to say, when we fix entropy density $\mathcal{S}$, the spherical symmetry case($k=1$) gives the upper bound of growth rate, which is larger than $4G_{N} \gamma^{\frac{1}{d}-1}z_{1}\mathcal{S}$ given by the planar symmetry case($k=0$). We also verify this conclusion by numerical calculation, which is shown in Figure \ref{Fignull8}.
\subsubsection{Fixed temperature}
We will calculate and compare the growth rate of different $k$ with fixed temperature in this subsection. The third line of (\ref{eq:405043}) implies that fixing the temperature $T$ is much more complicated since both $z_{h}$ and $f_{0}$ will be different with $k=1, 0, -1$. We can rewrite the temperature $T$ as
\begin{equation}
	\label{eq:4050435}
	T=\frac{d-2}{4\pi}\left(kz_{h}+\frac{z_{1}^{2}}{z_{h}}\right)\,,
\end{equation}
\begin{figure}
	\centering
	\includegraphics[width=0.45\textwidth]{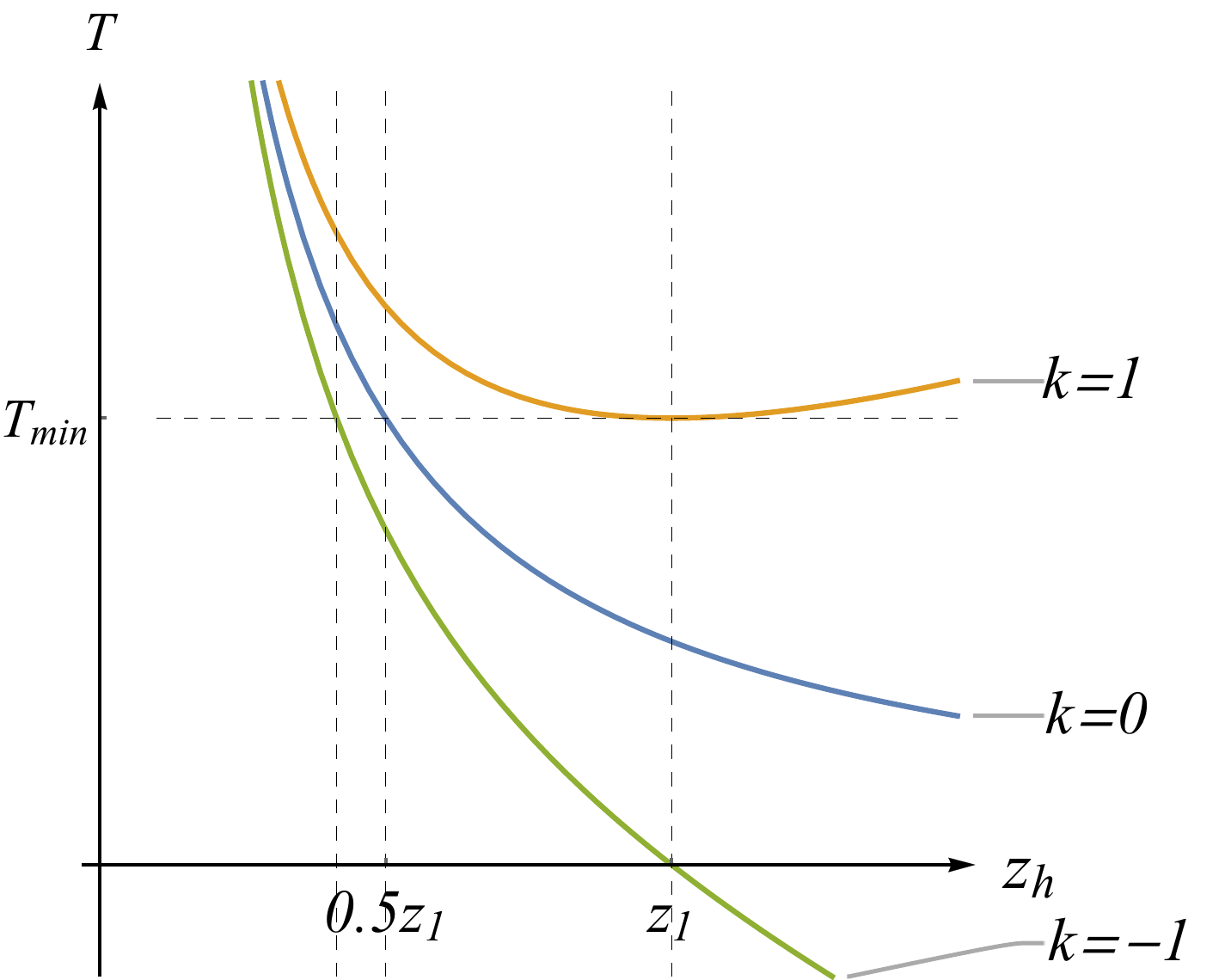}
	\caption{The sketch of $T(z_{h})$ with different $k$. The $T_{min}$ is the lower bound of the $f_{0}$.} \label{Fignull81}
\end{figure}
where $z_{1}$ is defined in \eqref{defz1}. The sketch of $T(z_{h})$ is shown as Figure \ref{Fignull81}. We can see that to keep the three cases have the same temperature $T$, there is a minimum $T_{min}$ of the temperature and the relationship of $z_{h}$ is $z_{h}(k=-1)<z_{h}(k=0)<z_{h}(k=1)$. There is an upper bound of spherical horizon $z_{h}(k=1)\leqslant z_{1}$. Similarly, there are also upper bounds $z_{1}/2$ and $(\sqrt{2}-1)z_{1}$ for the planar horizon and hyperbolic horizon respectively.
	
\begin{figure}
	\centering
	\subfigure[$d=3$]{
		\includegraphics[width=0.45\textwidth]{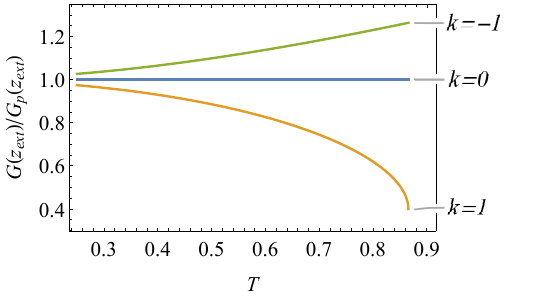}}\qquad
	\subfigure[$d=4$]{
		\includegraphics[width=0.45\textwidth]{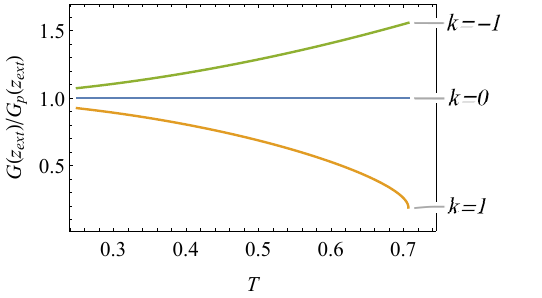}}

	\subfigure[$d=5$]{
		\includegraphics[width=0.45\textwidth]{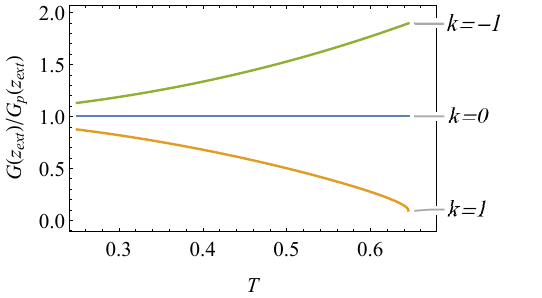}}\qquad
	\subfigure[$d=6$]{
		\includegraphics[width=0.45\textwidth]{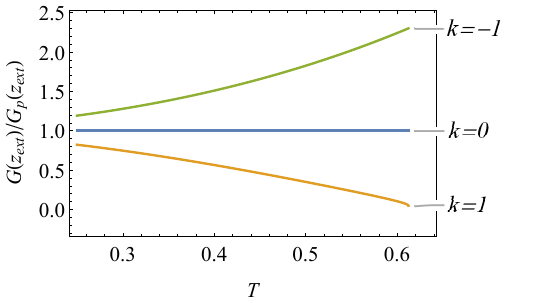}}
	\caption{For $\ell_{\rm AdS}=1$, the vertical axis represents $G(z_{\rm ext})/Gp(z_{\rm ext})$ and the horizontal axis represents $T$.} \label{Fignull9}
\end{figure}

For the planar case($k=0$), the maximum of $G(z)$ is again given by (\ref{eq:40506})
\begin{equation}
	\label{eq:4050436}
	G(z_{\rm ext})=\sqrt{\frac{d}{d-2}} \gamma^{\frac{1}{d}-1} \ell_{\rm AdS}^{-1} z_{h}^{1-d}=\left(\frac{d-2}{4\pi}\right)^{1-d} \gamma^{\frac{1}{d}-1}z_{1}^{3-2d}T^{d-1}.
\end{equation}
We can see that the growth rate of entanglement entropy is proportional to the $T^{d-1}$. When $k\neq 0$,  there are also no analytical expressions of $z_{\rm ext}$ and $G(z_{\rm ext})$. We adopt the prescription that calculating the derivative of $G(z_{\rm ext})$ with respect to a continuous real number $k$ in the interval $k \in [-1,1]$ again to compare $G(z_{\rm ext})$ with different $k$. We will first obtain (\ref{eq:4050433}), but the $\td f_{0}/\td k$ will be different. Since $z_{h}$ is no longer a constant, but a function of $k$. Combine the two conditions that $f(z_{h})=0$ and $\td T=0$, we finally obtain
\begin{equation}
	\label{eq:4050441}
	\frac{\td f_{0}}{\td k}=z_{h}^{2-d}+(2-d)z_{h}^{2-d}\frac{z_{1}^{2}+kz_{h}^{2}}{z_{1}^2-kz_{h}^{2}}.
\end{equation}
So the derivative of $G(z_{\rm ext})$ with respect to $k$ is
\begin{equation}
	\label{eq:4050442}
	\frac{\td G(z_{\rm ext})}{\td k}=\frac{z_{\rm ext}^{2-d}}{2G(z_{\rm ext})}\left[z_{h}^{2-d}\left(1+(2-d)\frac{z_{1}^{2}+kz_{h}^{2}}{z_{1}^2-kz_{h}^{2}}\right)-z_{\rm ext}^{2-d}\right].
\end{equation}
To compare the (\ref{eq:4050442}) and 0, we define a new function $\Theta(d,k,z_{h})$ as
\begin{equation}
	\label{eq:4050443}
	\Theta(d,k,z_{h})=z_{h}^{2-d}\left(1+(2-d)\frac{z_{1}^{2}+kz_{h}^{2}}{z_{1}^2-kz_{h}^{2}}\right)-z_{\rm ext}^{2-d}.
\end{equation}
Where $d\geqslant 3$, $-1\leqslant k \leqslant 1$, and $0 \textless z_{h} \leqslant (\sqrt{2}-1)z_{1}$. The $z_{\rm ext}$ is determined by
\begin{equation}
	\label{eq:4050444}
	kz_{h}^{2-d}+\frac{1}{\ell_{\rm AdS}^{2}z_{h}^{d}}=2kz_{\rm ext}^{2-d}+\frac{\gamma}{\ell_{\rm AdS}^{2}z_{\rm ext}^{d}}=f_{0}.
\end{equation}
When $k\geqslant 0$, there is obviously $\Theta(d,k,z_{h})\textless 0$. For the $k\textless 0$ case, we have to calculate $\Theta(d,k,z_{h})$ numerically. The result shows that $\Theta(d,k,z_{h})\textless 0$ always holds, which means $G(z_{\rm ext})$ is the decreasing function of $k$. That is to say, when we fix the temperature $T$, the hyperbolic symmetry case($k=-1$) gives the upper bound of growth rate, which is larger than $\left(\frac{d-2}{4\pi}\right)^{1-d} \gamma^{\frac{1}{d}-1}z_{1}^{3-2d}T^{d-1}$ given by the planar symmetry case($k=0$). We also verify this conclusion by numerical calculation, which is shown in Figure \ref{Fignull9}.

\subsection{Black holes with charge}
We have studied the maximum growth rates of holographic entanglement entropy for the vacuum black hole spacetime with different topologies. Now we will study the case of black holes with charge. We consider the simplest charged case, the RN-AdS black hole spacetime, which is also described by the metric (\ref{eq:4051}), and the function $f(z)$ and $\chi(z)$ is given by
\begin{equation}
	\label{eq:405087}
	f(z)=kz^{2}+\frac{1}{\ell_{\rm AdS}^{2}}-f_{0}z^{d}+\widetilde{q}z^{2d-2},\qquad \chi(z)=0
\end{equation}
Where the $f_{0}$ is the mass parameter, and $\widetilde{q}= q^{2}\geqslant 0$ is the charge parameter. Substitute (\ref{eq:405087}) into (\ref{eq:405044}), we obtain a similar equation as $(\ref{eq:405045})$
\begin{equation}
	\label{eq:4050871}
	2kz_{\rm ext}^{2}+\frac{2(d-1)}{(d-2)}\frac{1}{\ell_{\rm AdS}^{2}}-f_0z_{\rm ext}^{d}=0,
\end{equation}
The growth rate of charged case
\begin{equation}
	\label{eq:405088}
	G_{q}(z_{\rm ext})=\frac{\sqrt{-f(z_{\rm ext})}}{z_{\rm ext}^{d-1}}=\frac{\sqrt{-\left(kz^{2}+\frac{1}{\ell_{\rm AdS}^{2}}-f_0z^{d}+\widetilde{q}z^{2d-2}\right)}}{z_{\rm ext}^{d-1}}.
\end{equation}
Here we used subscript $q$ to imply that here we consider the charged cases.
	
Then the (\ref{eq:4050871}) will be identical to (\ref{eq:405045}), so will be the $z_{\rm ext}$. We have
\begin{equation}
	\label{eq:405090}
	\frac{\sqrt{-\left(kz^{2}+\frac{1}{\ell_{\rm AdS}^{2}}-f_{0}z^{d}+\widetilde{q}z^{2d-2}\right)}}{z_{\rm ext}^{d-1}} \leqslant \frac{\sqrt{-\left(kz^{2}+\frac{1}{\ell_{\rm AdS}^{2}}-f_{0}z^{d}\right)}}{z_{\rm ext}^{d-1}},
\end{equation}
since $\widetilde{q}\geqslant 0$, so we obtain
\begin{equation}
	\label{eq:405089}
	G_{q}(z_{\rm ext})\leqslant G_{0}(z_{\rm ext}).
\end{equation}
So the maximum growth rate of holographic entanglement entropy in the charged cases is less than that of the cases without charge when we fixed the mass parameter.
	
\begin{figure}
	\centering
	\subfigure[$d=3$]{
		\includegraphics[width=0.45\textwidth]{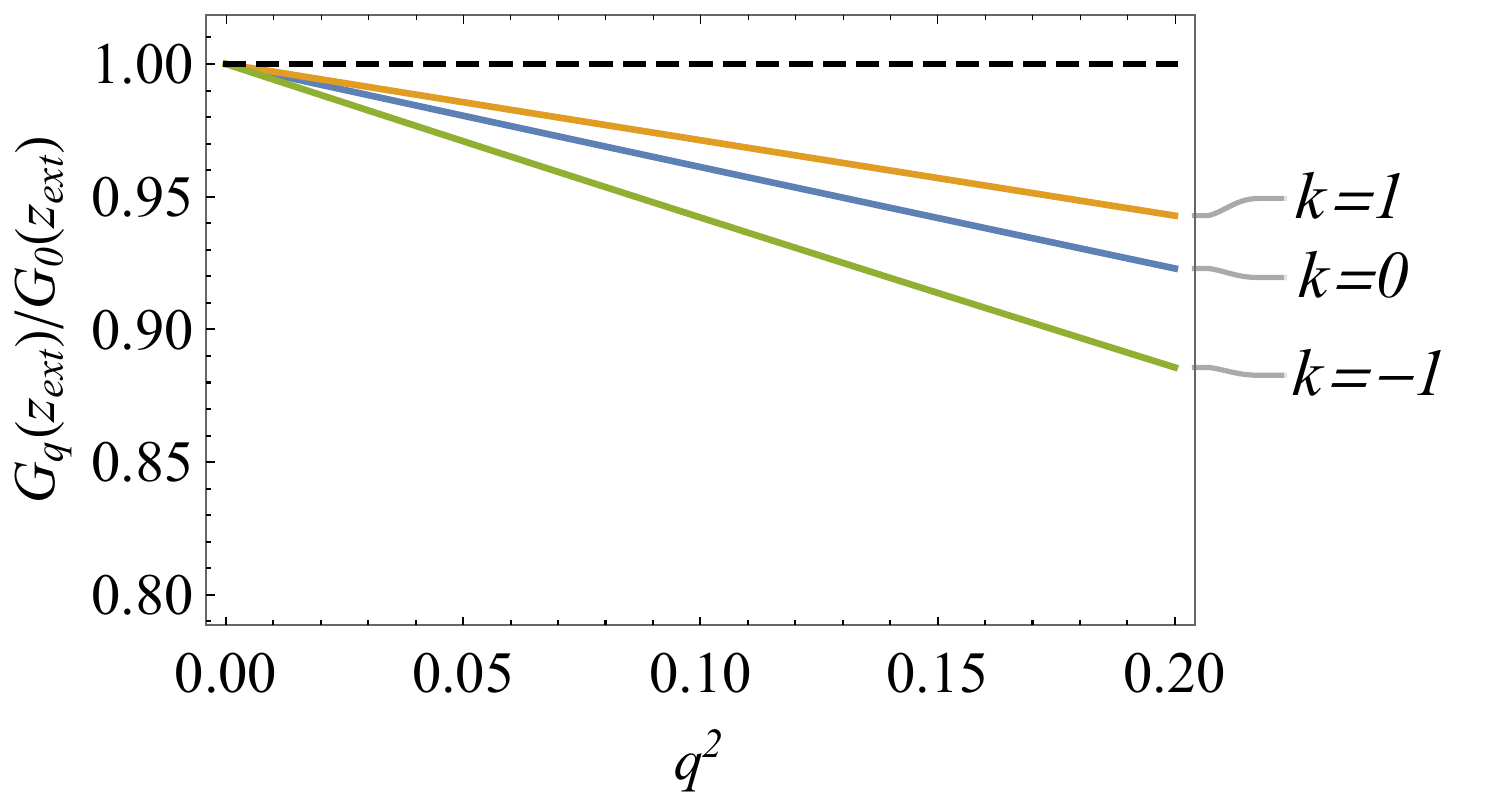}}\qquad
	\subfigure[$d=4$]{
		\includegraphics[width=0.45\textwidth]{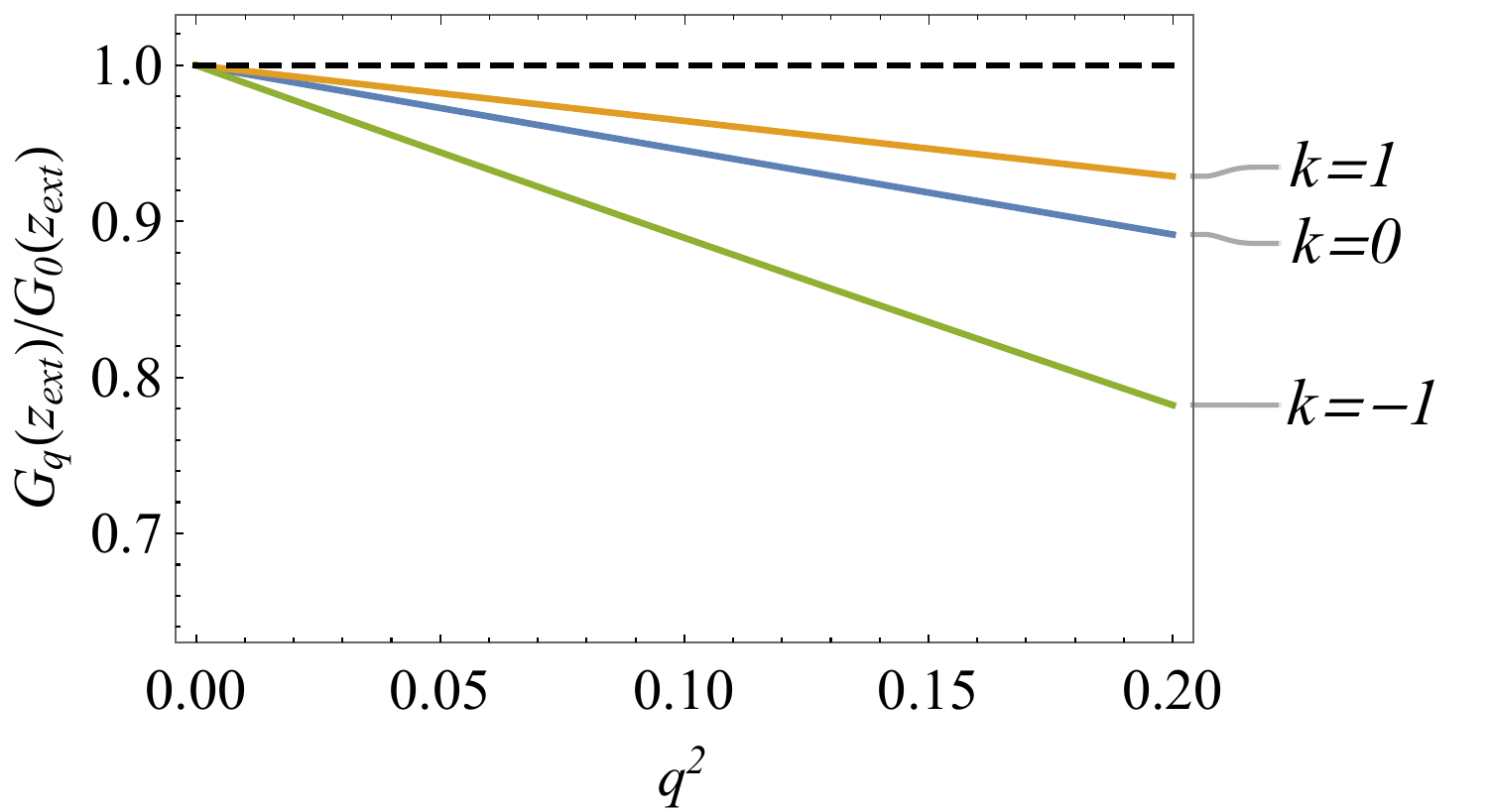}}

	\subfigure[$d=5$]{
		\includegraphics[width=0.45\textwidth]{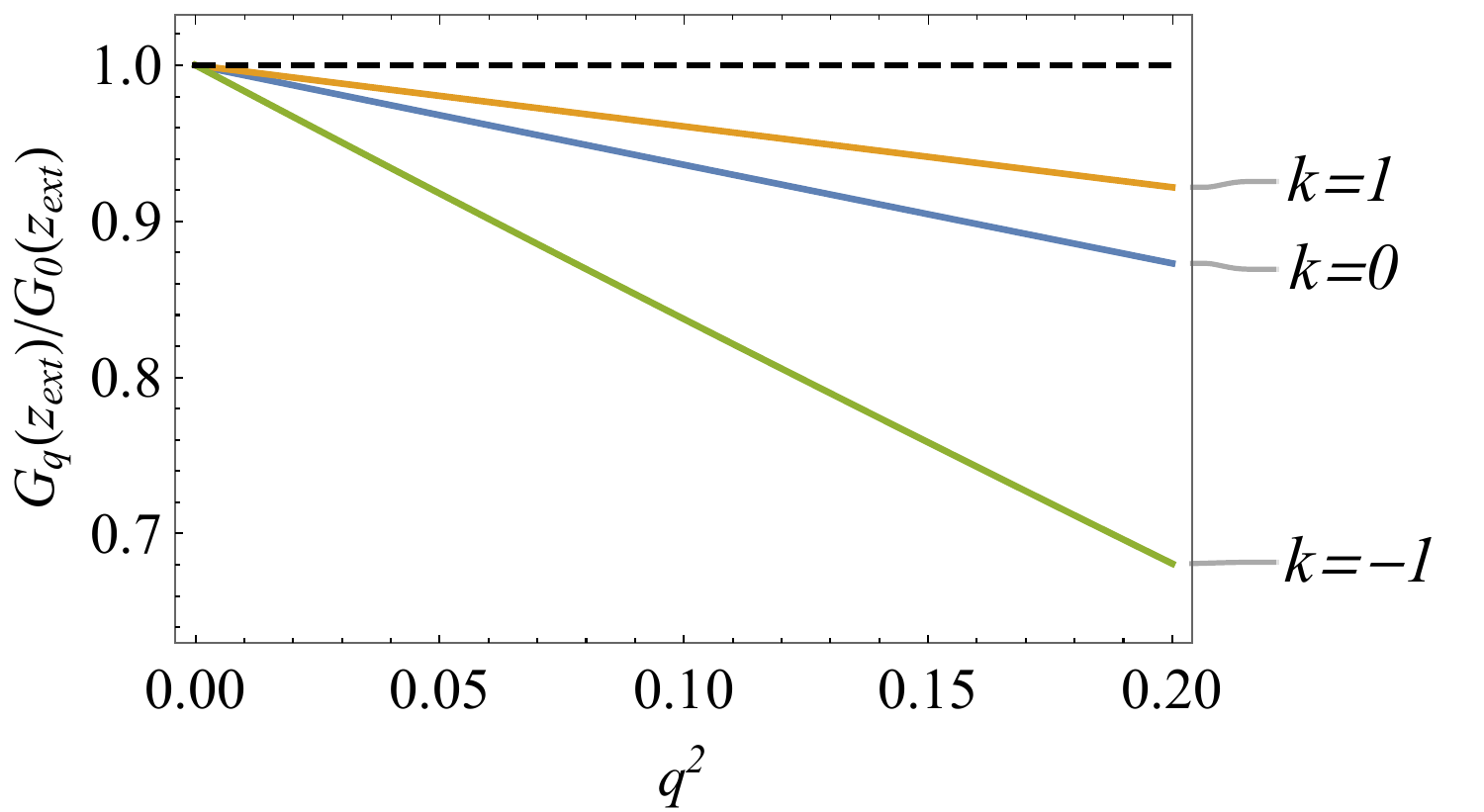}}\qquad
	\subfigure[$d=6$]{
		\includegraphics[width=0.45\textwidth]{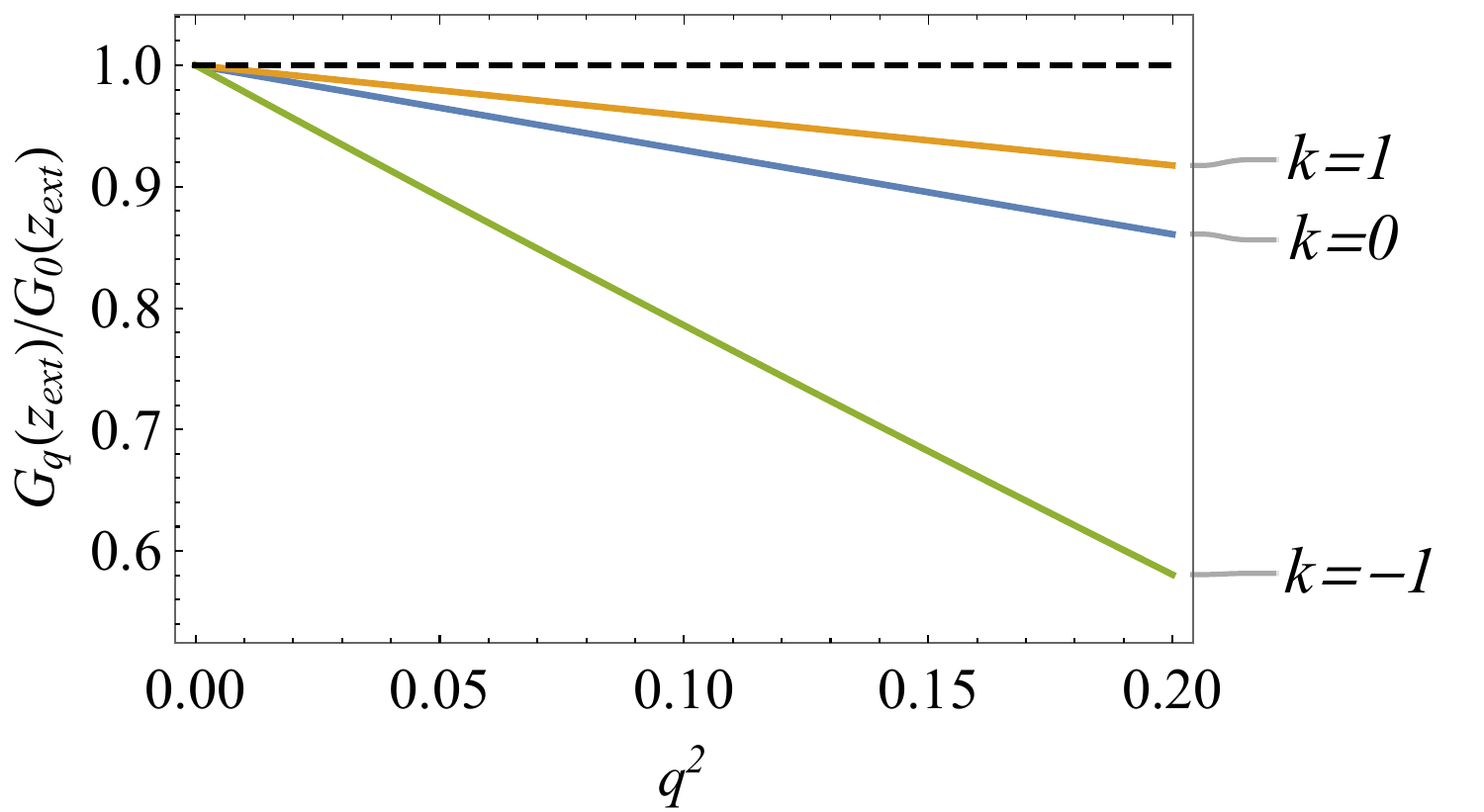}}
	\caption{For the case of $\ell_{\rm AdS}=1$ and fixed entropy density $\mathcal{S}=\frac{1}{4}$. The vertical axis represents $G_{q}(z_{\rm ext})/G_{0}(z_{\rm ext})$, where $G_{0}(z_{\rm ext})$ is the maximum growth rate of vacuum case.  The horizontal axis represents the charge parameter $q^{2}$.} \label{Fignull100}
\end{figure}
We next consider the cases that the entropy density $\mathcal{S}$ remains the same as that of vacuum black holes. According to the second line of (\ref{eq:405043}), the same $\mathcal{S}$ means the same $z_{h}$. For the charged cases, we choose the outermost horizon to calculate the entropy density. The numerical results are shown in the figure \ref{Fignull100}. We can see that the maximum growth rate will decrease as the charge parameter $\tilde{q}$ increases. So the growth rate in the charged cases is less than that of the cases without charge when we fixed the entropy density $\mathcal{S}$.
	
\begin{figure}
	\centering
	\subfigure[$d=3$]{
		\includegraphics[width=0.45\textwidth]{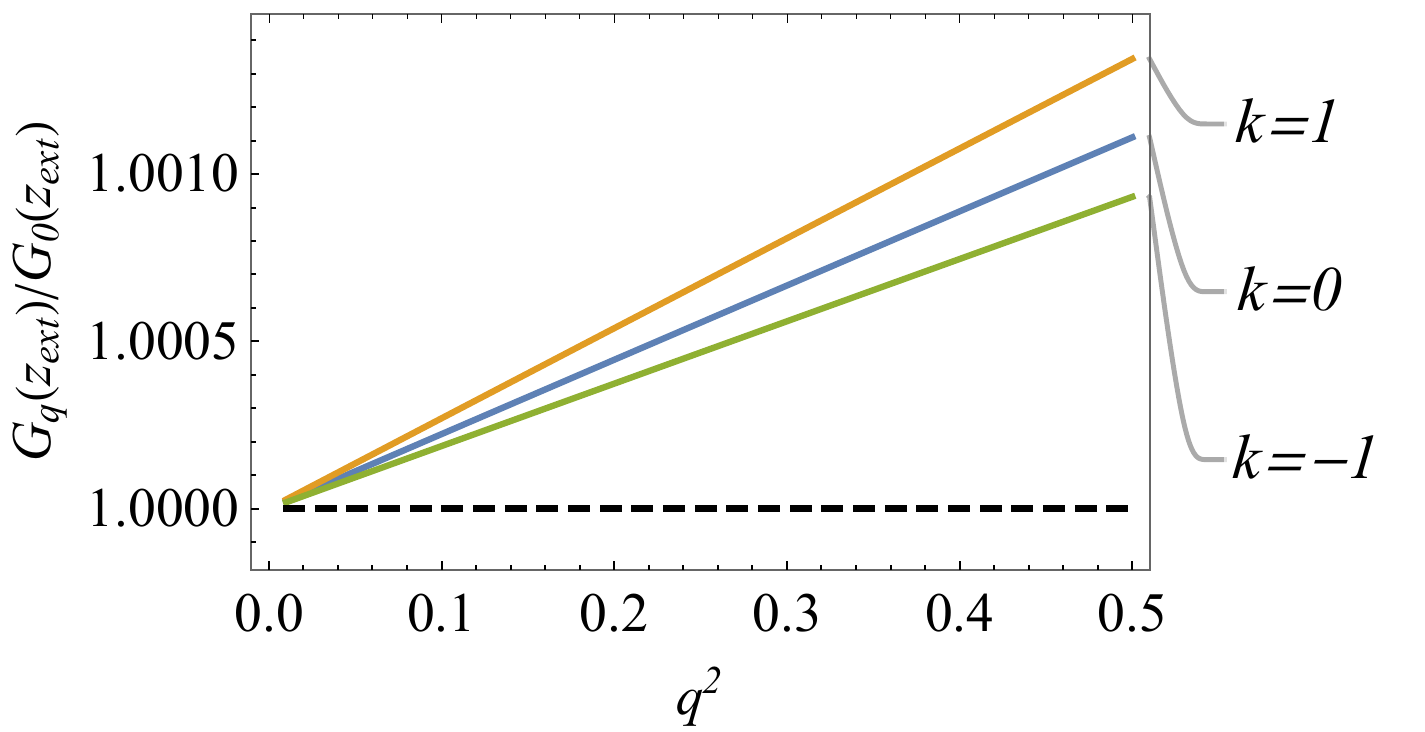}}\qquad
	\subfigure[$d=4$]{
		\includegraphics[width=0.45\textwidth]{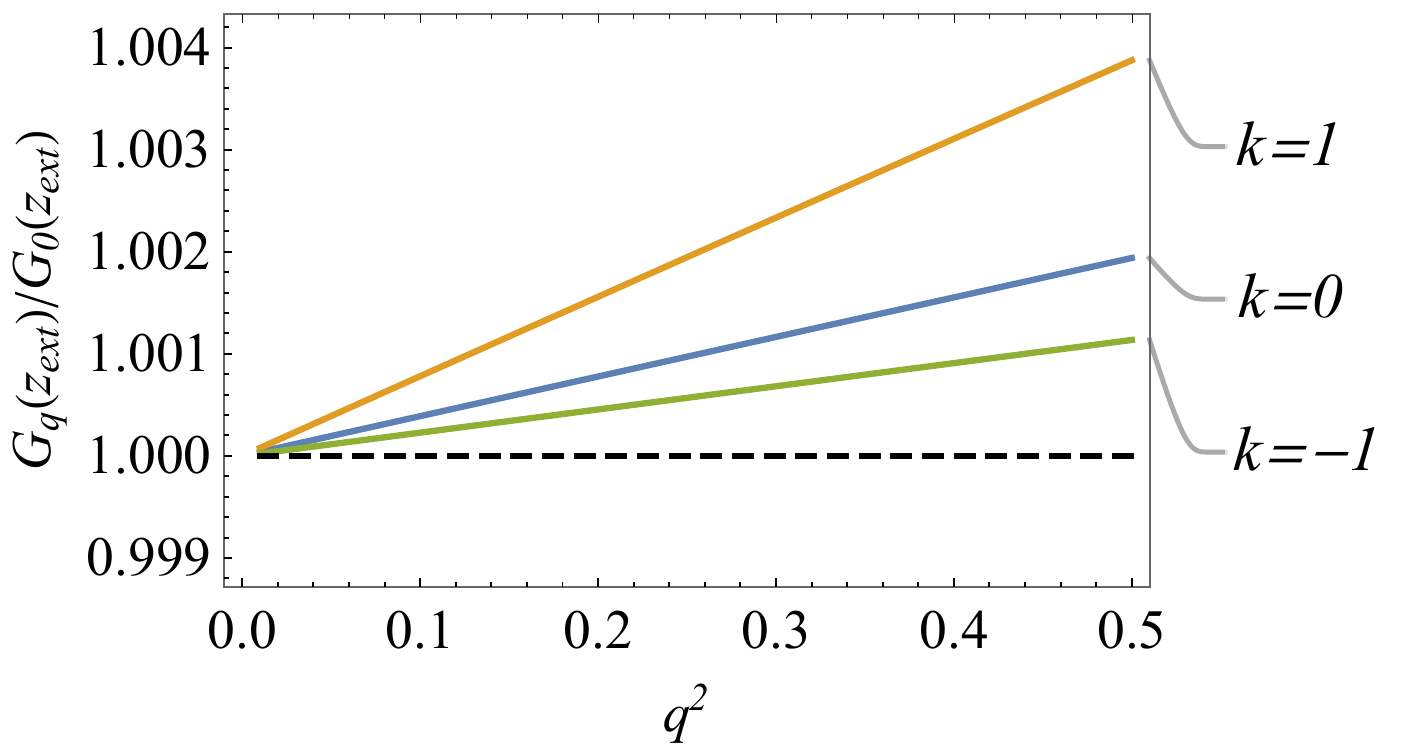}}

	\subfigure[$d=5$]{
		\includegraphics[width=0.45\textwidth]{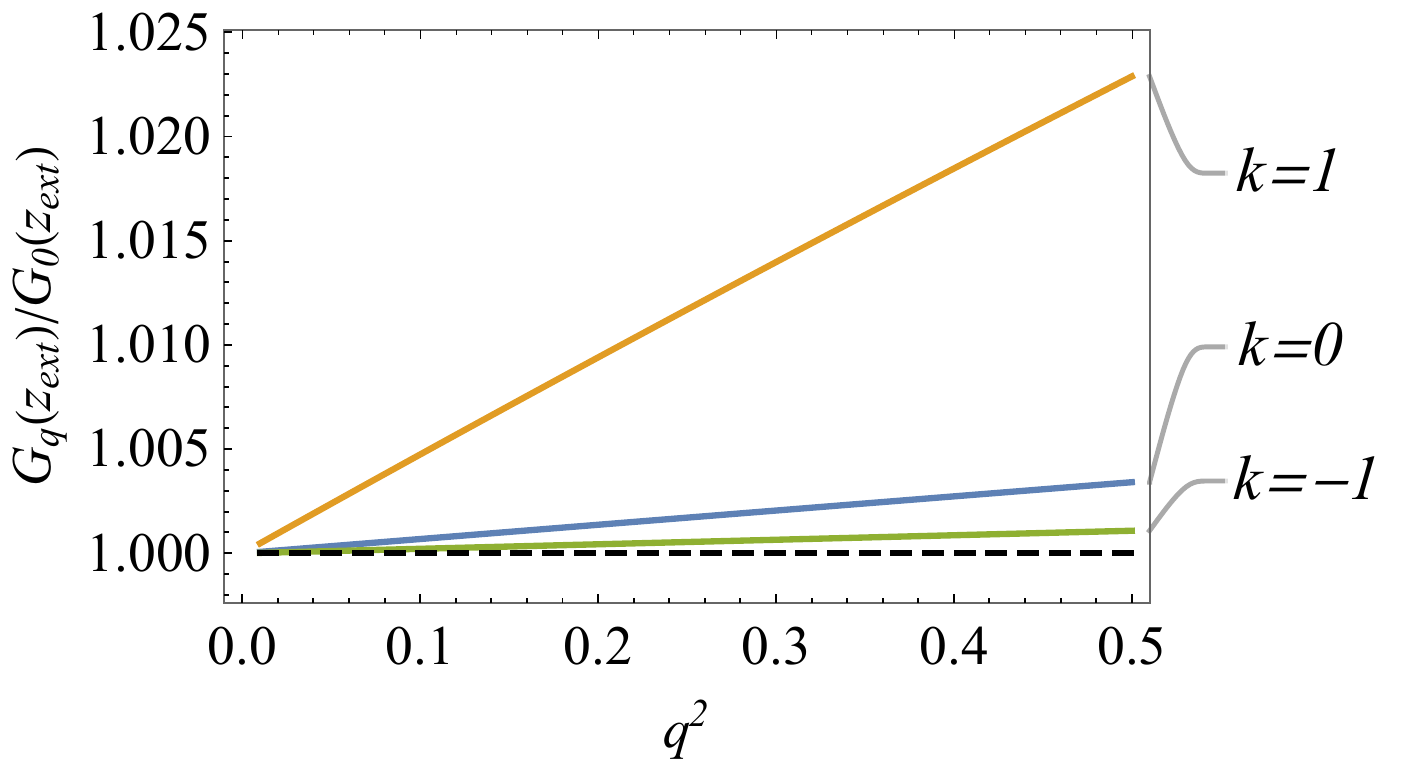}}\qquad
	\subfigure[$d=6$]{
		\includegraphics[width=0.45\textwidth]{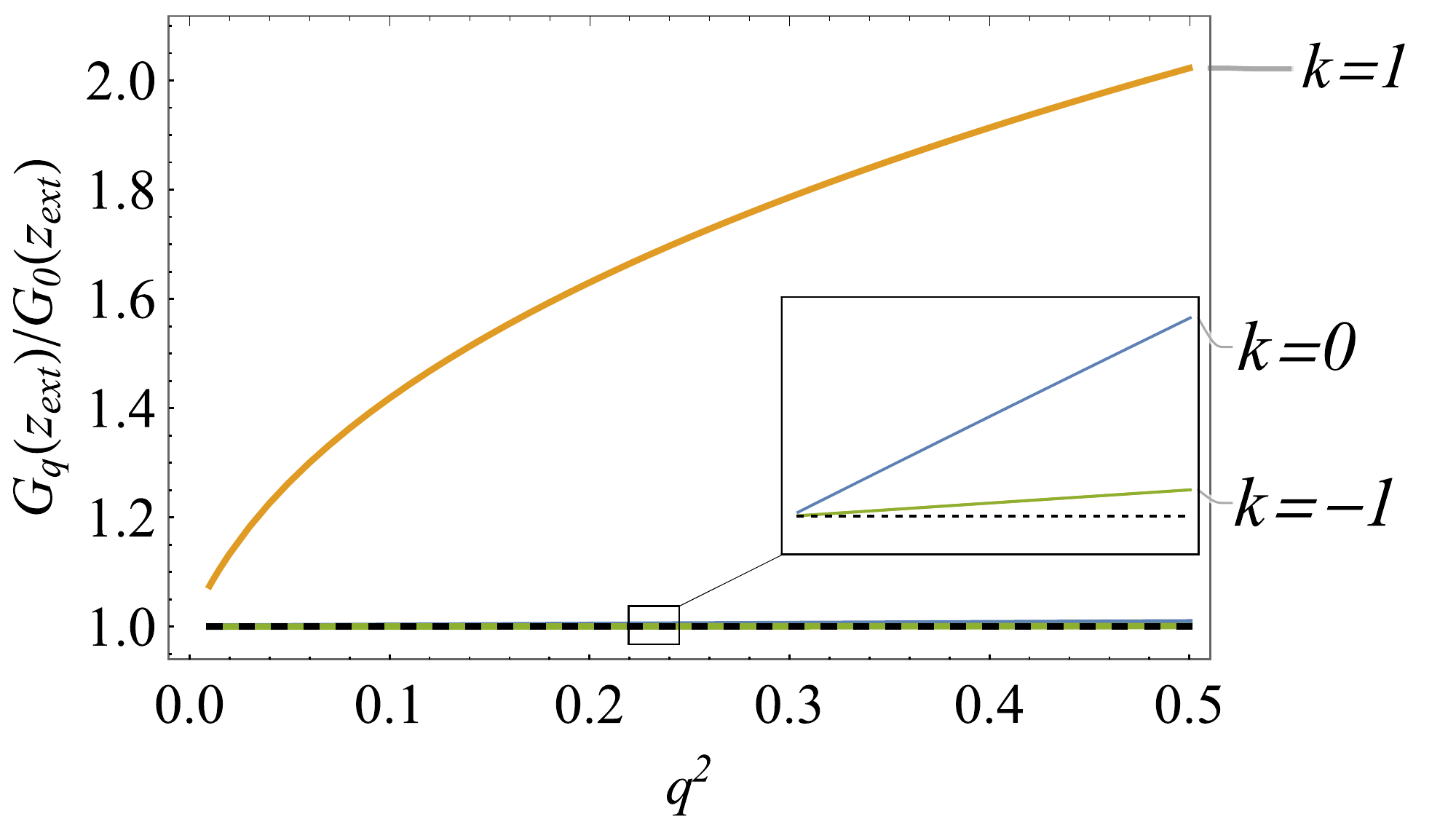}}
	\caption{For the case of $\ell_{\rm AdS}=1$ and fixed temperature $T=\frac{5}{2 \pi}$. the vertical axis represents $G_{q}(z_{\rm ext})/G_{0}(z_{\rm ext})$, where $G_{0}(z_{\rm ext})$ is the maximum growth rate of vacuum case.  The horizontal axis represents the charge parameter $q^2$.} \label{Fignull101}
\end{figure}

The cases with fixed temperature $T$ is different from other cases. The temperature is calculated by the third line of  (\ref{eq:405043}), and we also choose the outermost horizon to calculate the charged cases. The numerical results are shown in the figure \ref{Fignull101}. We can see that the maximum growth rate will increase as the charge parameter $\tilde{q}$ increases. So the growth rate in the charged cases is faster than that of the cases without charge when we fixed the temperature $T$. This conclusion is different from the previous case. The vacuum cases do not give the fastest growth rate when we fixed the temperature.
	
When we fix the mass and entropy density respectively, the vacuum cases give the faster growth rate.
Then we are going to ask, would the presence of other matter fields reduce the growth rate under the planar and spherical symmetries with fixed mass or entropy density? We conjecture: for all static planar or spherically symmetric asymptotically Schwarzschild-AdS black holes of same mass density or entropy density, the vacuum AdS black hole gives the maximum entanglement entropy growth rate. We will consider more general case and give proofs in next section.

\section{A proof about planar and spherically symmetric spacetime with matter fields}\label{section 4}
The results of the previous section suggest that, the maximum growth rate of giving energy and entropy density will decrease when there is a Maxwell field. In this section, we will show that such property is universal at least for spherically symmetric and planar symmetric asymptotically Schwarzschild-AdS black holes if matters satisfy the dominant energy condition\footnote{Asymptotically Schwarzschild-AdS spacetime requires that the asymptotic expansions of $f(z)$ and $\chi(z)$ in metric~\eqref{eq:4051} at AdS boundary $z=0$ are:
	\begin{equation}
		\nonumber
		\lim_{z \to 0}f(z)=kz^{2}+\frac{1}{\ell_{\rm AdS}^{2}}-f_{0}z^{d}+\mathcal{O}(z^{d+1}), \qquad
		\lim_{z \to 0}\chi(z)=\mathcal{O}(z^{d+1}).
	\end{equation}
When  matters do not decay rapidly enough, an asymptotically AdS spacetime may not be asymptotically Schwarzschild-AdS. For example, in our Sec.~\ref{section 5}, from Eq.~\eqref{eq:42049} we see that when the source term $\phi_\alpha$ is not zero, the spacetime is an asymptotically AdS spacetime but not asymptotically Schwarzschild-AdS. In  asymptotically Schwarzschild-AdS, the mass is determined completely by the bulk geometry but this is not true for general asymptotically AdS spacetime. From Eqs.~\eqref{eq:42052} and \eqref{eq:42055} we see that the mass in general an asymptotically AdS spacetime will also depend on the scheme of quantization. }. We here emphasize that the dominant energy condition is just one sufficient condition. In next section, we will give an example to show that such property can still be true even if the dominant energy condition is broken.

\subsection{Planar and spherical symmetry cases with fixed mass density}
In this subsection, we will keep the mass density to be fixed. And we use the metric (\ref{eq:410803}) to describe our spacetime with planar or spherical symmetry, which reads
\begin{equation}
	\label{eq:410803}
	\td s^{2} =\frac{1}{z^{2}}\left[-f \mathrm{e} ^{-\chi} \td t^{2}+\frac{\td z^2}{f}+ \mathfrak{s}_{i j}\td x^{i}\td x^{j}\right],
\end{equation}
where $f$ and $\chi$ are only functions of $z$, their expression is undetermined now. $\mathfrak{s}_{i j}$ is $d-2$ dimensional angular direction metric. The derivative of area with respect to time is given by (\ref{eq:405012}), and now we take its square
\begin{equation}
	\label{eq:410805}
	\dot{\mathcal A}^{2}= -4V_{k,d-2}^2 z^{2(1-d)} f\mathrm{e} ^{-\chi} .
\end{equation}
For the convenience of subsequent derivation in proving the inequality, we define $F(z) := -z^{2(1-d)} f \mathrm{e}^{-\chi/2}$, and then rewrite \eqref{eq:410805} as $\dot{\mathcal A}^{2} = 4V_{k,d-2}^{2} F(z) \mathrm{e}^{-\chi/2}$.

To facilitate the analysis, we cast the metric \eqref{eq:410803} into the Bondi–Sachs form by introducing the null coordinate
\begin{equation}
	u = t + \int 
	\frac{\mathrm{e}^{-\chi}}{f} \td z,
\end{equation}
and setting 
\begin{equation}
	\mathfrak{f} = f \mathrm{e}^{-\chi/2},\qquad \beta = -\frac{\chi}{4}.
\end{equation} 
Substitution yields exactly the Bondi–Sachs metric~\cite{Madler:2016xju}.
\begin{equation}
	\td s^2 = \frac{1}{z^2}\left[-\mathfrak{f} \mathrm{e}^{2\beta} \td u^2 + 2 \mathrm{e}^{2\beta} \td u \td z + \mathfrak{s}_{ij} \td x^i \td x^j\right].
\end{equation}
We substitute this metric into the Einstein equations and extract the $G_{zz}$ and $G_{uu}$ components, respectively. With $T_{\mu\nu}$ denoting the energy–momentum tensor of matter, we obtain the following equations
\begin{equation}
	\label{eq:4109}
	\partial_z \beta=-\frac{z \mathfrak{s}^{ik}\mathfrak{s}^{jl}(\partial_z \mathfrak{s}_{kl})(\partial_z \mathfrak{s}_{ij})}{8(d-1)}-\frac{4\pi z}{d-1}T_{zz},		
\end{equation}
and
\begin{equation}
	\label{eq:410901}
	-(d-1)z^{d-1}\partial_z(z^{-d}\mathfrak{f}) =\mathrm{e}^{2\beta}[\mathfrak{R}+2(\mathfrak{D}\beta)^2]-\mathfrak{D}^2 \mathrm{e}^{2\beta}+\frac{d(d-1)\mathrm{e}^{2\beta}}{z^2\ell_{\mathrm{AdS}}^2}-8\pi z^{-2}\mathrm{e}^{2\beta}(\rho-P).
\end{equation}
Here $\beta = - \chi/4$, $\mathfrak{f} = f \mathrm{e}^{-\chi/2}$. The quantity $T_{zz}$ is the $zz$ component of $T_{\mu\nu}$, and $\rho=T_{\mu\nu}n^{\mu}n^{\nu}$, $P=T_{\mu\nu}m^{\mu}m^{\nu}$, where $n^{\mu}$ and $m^{\mu}$ are orthogonal timelike and spacelike normal vectors of the subspace spanned by $\{x^{i}\}$. $\mathfrak{R}$ is the scalar curvature of the $(d-2)$-dimensional transverse space. For the planar or spherical symmetry cases, we can write $\mathfrak{R}$ as
\begin{equation}
	\mathfrak{R}=(d-1)(d-2)k, \label{eq:44311}
\end{equation}
with $k=0$ or $k=1$. $\mathfrak{D}$ is the covariant derivative with respect to $\mathfrak{s}_{ij}$. In the planar or spherically symmetric cases, the transverse metric $\mathfrak{s}_{ij}$ is independent of the radial coordinate $z$, and $\beta$ depends only on $z$. Therefore $\partial_z \mathfrak{s}_{ij} = 0$ and $\mathfrak D_i \beta = 0$. Equations \eqref{eq:4109} and \eqref{eq:410901} then reduce to
\begin{equation}
	\label{eq:410902}
	\partial_z \chi=\frac{16 \pi z}{d-1}T_{zz},
\end{equation}
and
\begin{equation}
	\label{eq:410903}
	-(d-1)z^{d-1}\partial_z(z^{-d}f \mathrm{e}^{-\frac{\chi}{2}}) =\mathrm{e}^{-\frac{\chi}{2}}(d-1)(d-2)k + \frac{d(d-1)\mathrm{e}^{-\frac{\chi}{2}}}{z^2\ell_{\mathrm{AdS}}^2}-8\pi z^{-2}\mathrm{e}^{-\frac{\chi}{2}}(\rho-P).
\end{equation}
The dominant energy condition implies $T_{zz} \geqslant 0$ and $\rho-P \geqslant 0$. Since $T_{zz} \geqslant 0$, Eq.~\eqref{eq:410902} gives
\begin{equation}
	\partial_{z} \chi \geqslant 0. \label{eq:4110}
\end{equation}
Together with the boundary condition $\chi|_{z=0}=0$, we get $\chi \geqslant 0$, which leads to $\mathrm{e}^{-\chi/2} \leqslant 1$. Combining this with \eqref{eq:410805} yields
\begin{equation}
	\dot{\mathcal A}^{2} \leqslant 4V_{k,d-2}^{2} F(z). \label{eq:411001}
\end{equation}

Next, we rewrite \eqref{eq:410903} as
\begin{equation}
	\frac{\td}{\td z}\left(z^{d-2} F\right)=\frac{d \mathrm{e} ^{-\chi}}{z^{d+1} \ell_{\mathrm{AdS}}^{2}} +
	\frac{ \mathrm{e} ^{-\chi}\mathfrak{R}}{(d-1) z^{d-1} }-\frac{ \mathrm{e} ^{-\chi}Q^{2}}{(d-1) z^{d-1} }, \label{eq:411002}
\end{equation}
where $Q=\sqrt{8 \pi (\rho-P)}/z\geq0$. Near the boundary $z\rightarrow0$, $\chi$ and $Q$ decay sufficiently fast, so we may neglect the term $- \mathrm{e} ^{-\chi}Q^{2}/((d-1) z^{d-1} )$ and integrate \eqref{eq:411002}
\begin{equation}
		\label{eq:4113}
		F \rightarrow -\frac{ k }{z^{d-2}}-\frac{1}{z^{d} \ell_{\mathrm{AdS}}^{2}}+f_{0},
\end{equation}
where $f_{0}$ is the integration constant. We can rewrite (\ref{eq:4113}) as
\begin{equation}
		\label{eq:411301}
		F(z) \rightarrow-z^{-2(1-d)}\left[kz^2+1 / \ell_{\mathrm{AdS}}^{2}-f_{0} z^{d}\right]=\left[G_{0}(z)\right]^{2}
\end{equation}
We find that $f_{0}$ can be regarded as the mass parameter which is fixed. And $G_{0}(z)$ is the growth rate function of the vacuum cases. According to (\ref{eq:411002}), for finite $z$, and consider that $\chi \geqslant 0$ and $k\geqslant0$, we have
\begin{equation}
		\label{eq:4114}
		\frac{\td}{\td z}\left(z^{d-2} F\right) \leqslant \frac{(d-2)k }{z^{d-1}}+\frac{d}{z^{d+1} \ell_{\mathrm{AdS}}^{2}}
\end{equation}
Integrating it and noting the asymptotically behavior (\ref{eq:411301}), we find
\begin{equation}
		\label{eq:4115}
		F(z) \leqslant \left[G_{0}(z)\right]^{2}
\end{equation}
Consider (\ref{eq:411001}), we finally get
\begin{equation}
	\label{eq:4116}
	\dot{\mathcal A}\leqslant 2V_{k,d-2} G_{0}(z)\leqslant 2V_{k,d-2} \max G_{0}.
\end{equation}
This result shows that, when the mass density is fixed, in the case of planar or spherically symmetric asymptotically Schwarzschild-AdS black holes with matter field that follow the dominant energy condition, the growth rate of holographic entanglement entropy will be less than the vacuum case.

\subsection{Planar and spherical symmetry cases with fixed entropy density}
The previous subsections considered the cases with fixed mass density. Now we consider the cases with fixed entropy density $\mathcal{S}$. Integrate the right side over $z$, we can write \eqref{eq:4114} as
\begin{equation}
	\label{eq:443142}
		\frac{\td}{\td z}\left(z^{d-2} F\right) \leqslant \frac{\td}{\td z}\left(-\frac{k}{z^{d-2}}-\frac{1}{z^{d} \ell_{\mathrm{AdS}}^{2}}+\tilde{f}_{0}\right)=\frac{\td}{\td z}\left(-\frac{f_{\text{sch}}(z)}{z^{d}}\right),
\end{equation}
where $\tilde{f}_{0}$ is the integration constant (independent of $z$) that satisfies
\begin{equation}\label{deffos1}
		k z_h^{2}+1 / \ell_{\mathrm{AdS}}^{2}-\tilde{f}_{0} z_h^{d}=0\,
\end{equation}
and we define $f_{\text{sch}}(z):=k z^{2}+1 / \ell_{\mathrm{AdS}}^{2}-\tilde{f}_{0} z^{d}$. According to the second line of (\ref{eq:405043}), the fixed $\mathcal{S}$ means the fixed horizon coordinate $z_{h}$. Since the metric is (\ref{deffos1}), we have $f_{\text{sch}}(z_{h})=F(z_{h})=0$. This shows that
\begin{equation}
	\label{eq:443143}
	z_{h}^{d-2} F(z_{h}) = -\frac{f_{\text{sch}}(z_{h})}{z_{h}^{d}}=0.
\end{equation}
When $z>z_{h}$, combine (\ref{eq:443142}) and (\ref{eq:443143}), we have
\begin{equation}
	\label{eq:443144}
	z^{d-2} F(z) \leqslant -\frac{f_{\text{sch}}(z)}{z^{d}}.
\end{equation}
So we obtain
\begin{equation}
	\label{eq:443145}
	F(z) \leqslant -\frac{f_{\text{sch}}(z)}{z^{2d-2}}= \left[G_{0}(z)\right]^{2},
\end{equation}
where $G_{0}(z)$ is the growth rate function of the vacuum black hole cases. We finally get
\begin{equation}
	\label{eq:4431002}
	\dot{\mathcal A}\leqslant 2 V_{k,d-2} G_{0}(z)\leqslant 2V_{k,d-2} \max G_{0}.
\end{equation}
This result shows that, when the entropy density is fixed, in the case of planar or spherically symmetric asymptotically Schwarzschild-AdS black holes with matter field that follow the dominant energy condition, the growth rate of holographic entanglement entropy will less than the vacuum case.

In the case of RN black hole we have found that the Maxwell field always decreases the maximal growth rate of entanglement entropy. However, the proofs here only cover the spherically and planar symmetric cases. The reason is that we require $\mathfrak{R}\geqslant0$ to obtain inequality~\eqref{eq:4114} from Eq.~\eqref{eq:4110}. It is not clear that whether dominant energy condition can insure the same conclusion or not for hyperbolically symmetric black hole. We hope we could address this issue in the future.

\section{Black holes with scalar hair}\label{section 5}
We have proved that the vacuum black holes with $k\geqslant0$ have the maximum growth rate by the dominant energy condition and the Einstein equation.
In this section, we consider the planar symmetric AdS black holes with real scalar hair. We will choose the negative ``mass-square'' for the scalar field so that the dominant energy condition can be broken. In addition, due to the presence of external source, the bulk spacetime is asymptotically AdS but not asymptotically Schwarzschild-AdS.
	
The model is Einstein gravity with a negative cosmological minimally coupled to a real scalar field $\phi$, whose action reads
\begin{equation}
	\label{eq:4201}
	S=\frac{1}{16\pi}\int \mathrm{d}^{d+1} x \sqrt{-g}\left(R-2\Lambda-\frac{1}{2} \nabla^{\mu} \phi \nabla_{\mu} \phi-\frac{1}{2}m^2\phi^2\right),
\end{equation}
where $\Lambda=-\frac{d(d-1)}{2\ell_{\rm AdS}^{2}}$, and $m$ is the mass parameter of the scalar field $\phi$. Performing variation on the action (\ref{eq:4201}) with respect to the metric $g_{\mu\nu}$, and the scalar field $\phi$, we obtain the equation of motion
\begin{equation}
		\label{eq:4202}
		\begin{aligned}
			R_{\mu \nu}-\frac{1}{2} R g_{\mu \nu}+\Lambda g_{\mu \nu}&=\frac{1}{2}\left( \nabla_{\mu} \phi \nabla_{\nu}\phi -\frac{1}{2} g_{\mu \nu} \nabla_{\mu} \phi \nabla^{\mu} \phi- \frac{1}{2}g_{\mu \nu}m^2\phi^2\right),\\
			\nabla_{\mu} \nabla^{\mu} \phi-m^2\phi&=0.
		\end{aligned}
\end{equation}
Consider the planar symmetry ansatz given by
\begin{equation}
		\label{eq:4203}
		\begin{aligned}
			d s^{2}&=-f(r) e^{-\chi(r)} \td t^{2}+\frac{\td r^{2}}{f(r)}+r^{2} \td \mathbf{x}_{d-1}^2, \\
			\phi &= \phi (r).
		\end{aligned}
\end{equation}
where $f, \chi, \phi$ are only functions of coordinate $r$, and $\td \mathbf{x}_{d-1}^2=\td x_{1}^{2}+\td x_{2}^{2}+\cdots+\td x_{d-1}^{2}$ is the $d-1$ dimensional spatial directional line element.

With the ansatz, the equation of motion (\ref{eq:4202}) reduce to
\begin{equation}
		\label{eq:4204}
		\begin{aligned}
			\frac{\chi^{\prime}}{r}+\frac{1}{d-1} \phi^{\prime 2}&=0, \\
			\frac{2}{r} \frac{f^{\prime}}{f}-\frac{\chi^{\prime}}{r}+\frac{1}{d-1} \frac{m^2\phi^2}{f}+\frac{2(d-2)}{r^{2}}&=0, \\
			\frac{f^{\prime \prime}}{f}-\chi^{\prime \prime}+\frac{1}{2} \chi^{\prime 2}+\frac{(d-2)\chi^{\prime}}{r}+\left(\frac{d-3}{r}-\frac{3}{2} \chi^{\prime}\right) \frac{f^{\prime}}{f}-\frac{2(d-2)}{r^{2}}&=0,\\
			\phi^{\prime \prime}+\left(\frac{f^{\prime}}{f}-\frac{\chi^{\prime}}{2}+\frac{d-1}{r}\right) \phi^{\prime}-\frac{m^2\phi}{f} &=0.
		\end{aligned}
\end{equation}
Since there are only three independent functions of coordinate $r$, only three equations of (\ref{eq:4204}) are independent.\footnote{One can verify that by transforming the first three equations to get the fourth equation.} For convenience, we can choose the first, second and fourth equations for later calculation.

The entropy density and temperature are given by
	
\begin{equation}
		\label{eq:42041}
		\mathcal{S}=\frac{1}{4}r_{h}^{d-1}, \qquad
		T=\frac{1}{4 \pi}f^{\prime}(r_{h}) {\rm e}^{-\frac{\chi(r_{h})}{2}}.
\end{equation}
	
We assume that the solution is an asymptotically AdS spacetime, which means the functions $f$, $\chi$, $\phi$ have the following behavior at the AdS boundary
\begin{equation}
		\label{eq:42042}
		\lim_{r \to \infty}f(r)=\frac{r^2}{\ell_{\rm AdS}^{2}}, \qquad
		\lim_{r \to \infty}\chi(r)=0, \qquad
		\lim_{r \to \infty}\phi(r)=0.
\end{equation}
	The scalar field $\phi(r)$ can be expanded at the AdS boundary($r\rightarrow\infty$) as~\cite{Skenderis:2002wp,Marolf:2006nd}
\begin{equation}
		\label{eq:42044}
		\phi(r)=\frac{\phi_{\alpha}}{r^{d-\Delta}}\left(1+\cdots\right)+\frac{\phi_{\beta}}{r^{\Delta}}\left(1+\cdots\right),
\end{equation}
where $\phi_{\alpha}$ and $\phi_{\beta}$ are non-trivial parameters. According to the holographic renormalization of the massive scalar field~\cite{Skenderis:2002wp,Marolf:2006nd}, the parameter $\Delta$, which is called the scaling dimension of the boundary field, is given by
\begin{equation}
		\label{eq:42045}
		\Delta=\frac{1}{2}\left(d + \sqrt{d^2+4m^2\ell_{\rm AdS}^{2}}\right),
\end{equation}
where $m$ is the mass of the bulk scalar field. The surd in (\ref{eq:42045}) implies that $m^2$ can be negative, and there is a lower bound of $m^2$ which is called the Breitenlohner-Freedman (BF) bound $m^2_{BF}=-\frac{d^2}{4\ell_{\rm AdS}^{2}}$~\cite{Breitenlohner:1982jf,Breitenlohner:1982bm}. Hence, the scalar field may not obey the dominant energy conditions.
	
Now we consider a specific case to calculate the growth rate of holographic entanglement entropy. Without loss of generality, we will set $\ell_{\rm AdS}=1$ in this section. We take four-dimensional space-time$(d=3)$, and choose the mass parameter  $m^2=-2$.\footnote{Note that it is satisfied the Breitenlohner-Freedman bound, $m^2=-2 \textgreater m^2_{BF}=-2.25$.} And we can get the scaling dimension $\Delta = 2$ by (\ref{eq:42045}). Substitute it into (\ref{eq:42044}), we obtain asymptotic expansion of $\phi(r)$ on the AdS boundary
\begin{equation}
		\label{eq:42047}
		\phi(r)=\frac{\phi_{\alpha}}{r}+\frac{\phi_{\beta}}{r^{2}}+\frac{\phi_3}{r^{3}}+\mathcal{O}(\frac{1}{r^{4}}),
\end{equation}
where $\phi_3$ is undetermined coefficient that can be represented by $\phi_{\alpha}$, $\phi_{\beta}$. Since here we choose negative mass-square, the dominant energy condition may be broken somewhere in the bulk. We can also write the expansion of $f(r)$ and $\chi(r)$ on the AdS boundary according to (\ref{eq:42042})
\begin{equation}
	\label{eq:42048}
	\begin{aligned}
		f(r)&=r^2\left[1+\frac{f_2}{r^{2}}+\frac{f_3}{r^{3}}+\mathcal{O}(\frac{1}{r^{4}})\right], \\
		\chi(r)&=\frac{\chi_1}{r}+\frac{\chi_2}{r^{2}}+\frac{\chi_3}{r^{3}}+\mathcal{O}(\frac{1}{r^{4}}),
	\end{aligned}
\end{equation}
with undetermined parameters $\{f_2,f_3, \chi_1,\chi_2,\chi_3\}$.
Substitute (\ref{eq:42047}) and (\ref{eq:42048}) into the equations  (\ref{eq:4204}) at $r\rightarrow \infty$, we obtain~\cite{Li:2020spf}
\begin{equation}
		\label{eq:42049}
		\begin{aligned}
			\phi(r)&=\frac{\phi_{\alpha}}{r}+\frac{\phi_{\beta}}{r^{2}}-\frac{\phi_{\alpha}^3}{8r^{3}}+\mathcal{O}(\frac{1}{r^{4}}),\\
			f(r)&=r^2\left[1+\frac{\phi_{\alpha}^2}{4 r^{2}}+\frac{f_3}{r^{3}}+\mathcal{O}(\frac{1}{r^{4}})\right], \\
			\chi(r)&=\frac{\phi_{\alpha}^2}{4 r^{2}}+\frac{2\phi_{\alpha}\phi_{\beta}}{3r^{3}}+\mathcal{O}(\frac{1}{r^{4}}),
		\end{aligned}
\end{equation}
there are only three free parameters $\phi_{\alpha}$, $\phi_{\beta}$ and $f_3$, which will give us the energy(or the mass) of space-time. In fact, according to the boundary conditions of the equation (\ref{eq:4204}), we can find that $\phi_{\alpha}$ and $\phi_{\beta}$ are related, that is, there are only two free parameters. One can see that the function $f(r)$ is not asymptotically Schwarzschild if the coefficient $\phi_{\alpha,\beta}\neq0$.

In order to obtain the total energy, we need the prescription of holographic renormalization. Since there are two parameters $\phi_{\alpha}$ and $\phi_{\beta}$ in the ~\eqref{eq:42044}, we can fix them respectively at the boundary. This will give us two different quantization schemes~\cite{Marolf:2006nd}, and we will discuss them reseparately.

We first consider standard quantization scheme which takes the leading order $\phi_{\alpha}$ as the source,  the boundary action will given by~\cite{Marolf:2006nd}
\begin{equation}
		\label{eq:42050}
		S_{\partial}^{(\alpha)}=\frac{1}{16 \pi} \int_{r \rightarrow \infty} \td x^{3} \sqrt{-h}\left[2 K-4-\frac{1}{2} \phi^{2}\right]
\end{equation}
where $h=\rm{det}$$ (h_{ij})$, $h_{ij}$ is the induced metric at the AdS boundary, and $K$ is the extrinsic curvature scalar of the boundary manifold. The first term of (\ref{eq:42050}) is the Gibbons-Hawking-York term. The second and third terms are counter terms for removing divergences.\footnote{In most cases, there is a finite counter term (the contact term) $-c \phi^{3}$ after the counter terms in (\ref{eq:42050})~\cite{Skenderis:2002wp,Li:2020spf}. Here we choose the minimum coupling case with $c=0$.}
	
Then we can obtain the holographic stress tensor by performing variation on the boundary action (\ref{eq:42050})
\begin{equation}
		\label{eq:42051}
		T_{\mu \nu}^{(\alpha)}=\frac{1}{16\pi} \lim _{r \rightarrow \infty} r\left[2\left(K h_{\mu \nu}-K_{\mu \nu}-2 h_{\mu \nu}\right)-\frac{1}{2}h_{\mu \nu} \phi^{2} \right].
\end{equation}
Substituting the expansions (\ref{eq:42049}) into the stress tensor (\ref{eq:42051}), we obtain the $T_{tt}$ component reads
\begin{equation}
		\label{eq:42052}
		16 \pi T_{t t}^{(\alpha)}=16 \pi  \mathcal{E}^{(\alpha)}=-2 f_3+\phi_{\alpha} \phi_{\beta},
\end{equation}
where $\mathcal{E}^{(\alpha)}$ is the energy(or mass) density. When the parameter $\phi_{\alpha} = 0$, the scalar field will vanish, the metric will degenerate to vacuum black hole solution.
	
Now we solve the equation (\ref{eq:4204}) numerically. The asymptotic infinity of the numerical solution will give the parameters $\phi_{\alpha}$, $\phi_{\beta}$ and $f_3$, and then we can compute the energy $\mathcal{E}^{(\alpha)}$ by (\ref{eq:42052}). We are interested in different cases that with the same energy. So we vary the parameters $\phi_{\alpha}$and $f_3$ while keeping the $\mathcal{E}^{(\alpha)}$ a constant.
	
\begin{figure}
	\centering
	\subfigure[$\mathcal{E}^{(\alpha)}=1/16 \pi$]{
		\includegraphics[width=0.4\textwidth]{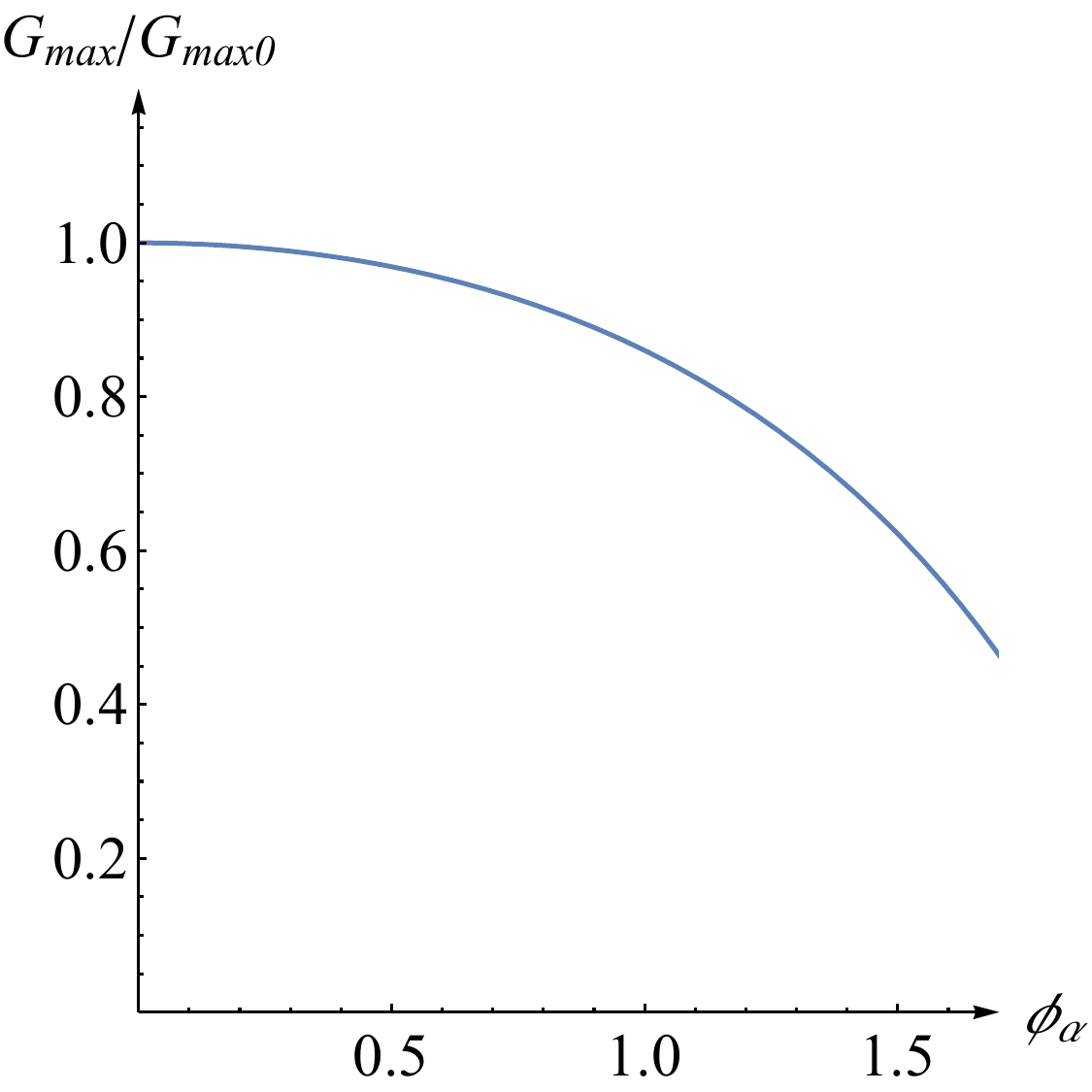}}\qquad
	\subfigure[$\mathcal{S}=0.25 $]{
		\includegraphics[width=0.4\textwidth]{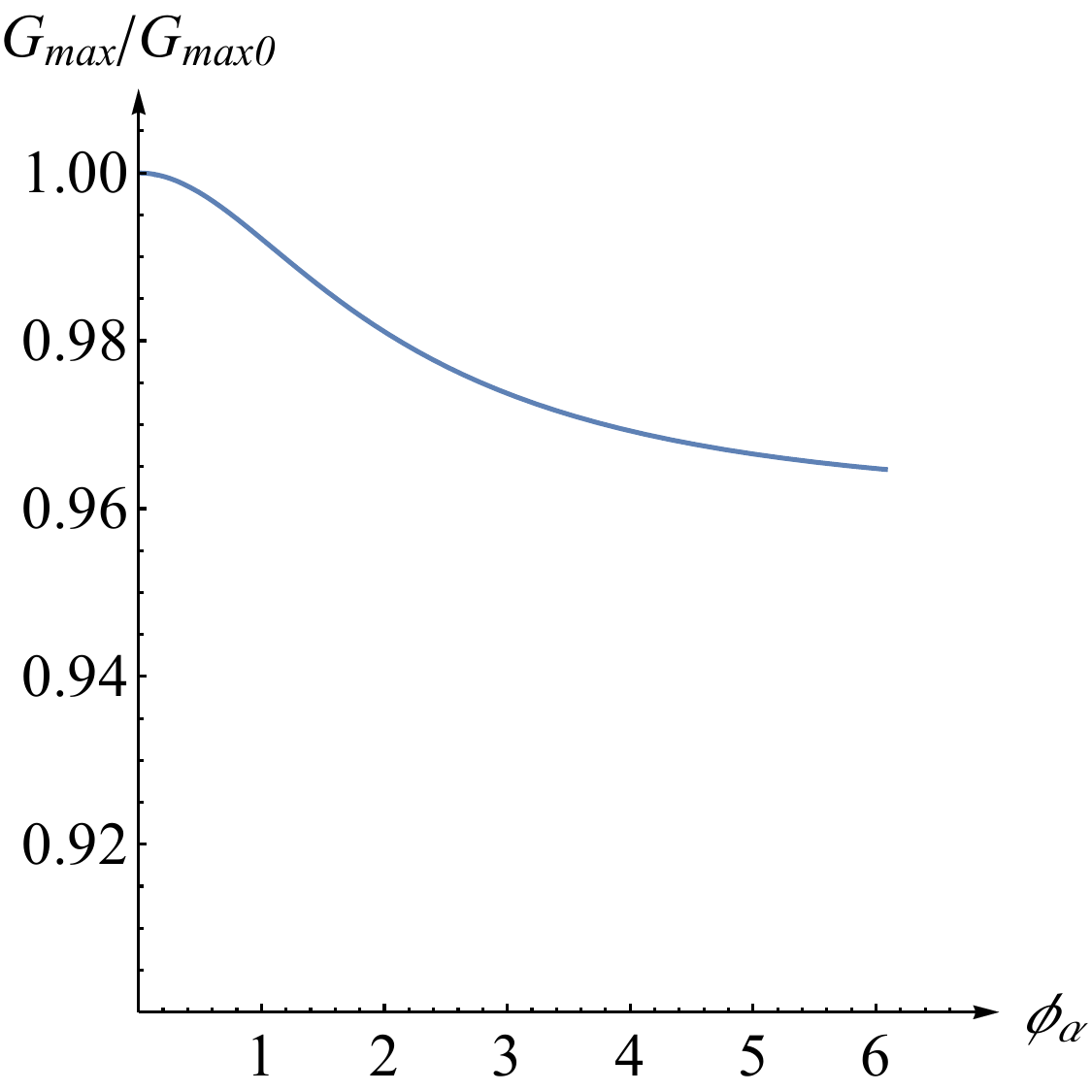}}\qquad
	\caption{For $\ell_{\rm AdS}=1$, relationship between growth rate and $\phi_{\alpha}$. Fixed the $\mathcal{E}=1/16 \pi$,  $\mathcal{S}=0.25 $ respectively, the larger the parameter $\phi_{\alpha}$, the smaller the maximum value of the growth rate function $G(r)$. } \label{Fignull99}
\end{figure}

We can compute the growth rate of holographic entanglement entropy by (\ref{eq:40593}). Figure \ref{Fignull99}(a) shows the maximum value of growth rate function with different $\phi_{\alpha}$. When the parameter $\phi_{\alpha} = 0$, we have the vacuum solution, which has the maximum growth rate. When we increase the parameter $\phi_{\alpha}$, which means a stronger scalar field in spacetime, the maximum growth rate will decrease. That is to say, vacuum black holes will have a greater growth rate of holographic entanglement entropy.
	
We next consider black holes with the same entropy density $\mathcal{S}$ but different scalar fields, which means we have a fixed horizon radius $r_{h}$. We change the parameter of the scalar field, which will give us the growth rate function curves(see Figure \ref{Fignull99}(b)). It still follows the law that vacuum black holes have the greater growth rate, and increasing the scalar field reduces the growth rate.

We next consider alternative quantization scheme which takes the sub-leading order $\phi_{\beta}$ as the source\footnote{Noting that $\phi_{\alpha}$ and $\phi_{\beta}$ are related. If $\phi_{\beta}$ is positive, $\phi_{\alpha}$ will be negative. Since $\phi_{\alpha}$ and $\phi_{\beta}$ always appear in the energy expression in the form of $\phi_{\alpha}\phi_{\beta}$, there is a symmetry of the $\phi_{\beta}$ and $-\phi_{\beta}$. Here we set $\phi_{\beta}$ is positive.}, the boundary action is given by~\cite{Marolf:2006nd}
\begin{equation}
	\label{eq:42053}
	S_{\partial}^{(\beta)}=\frac{1}{16 \pi} \int_{r \rightarrow \infty} \td x^{3} \sqrt{-h}\left[2 K-4+\phi n^{a} \partial_{a}\phi +\frac{1}{2} \phi^{2}\right],
\end{equation}
where the $n^{a}$ is the outward unit normal vector of the AdS boundary. Then we can obtain the holographic stress tensor as
\begin{equation}
	\label{eq:42054}
	T_{\mu \nu}^{(\beta)}=\frac{1}{16\pi} \lim _{r \rightarrow \infty} r\left[2\left(K h_{\mu \nu}-K_{\mu \nu}-2 h_{\mu \nu}\right)+h_{\mu \nu}(\phi n^{a} \partial_{a}\phi +\frac{1}{2} \phi^{2}) \right].
\end{equation}
The total energy is given by
\begin{equation}
	\label{eq:42055}
	16 \pi  \mathcal{E}^{(\beta)}=16 \pi T_{t t}^{(\beta)}=-2 f_3+2\phi_{\alpha} \phi_{\beta}.
\end{equation}

\begin{figure}
	\centering
	\subfigure[$\mathcal{E}^{(\beta)}=1/16 \pi$]{
		\includegraphics[width=0.4\textwidth]{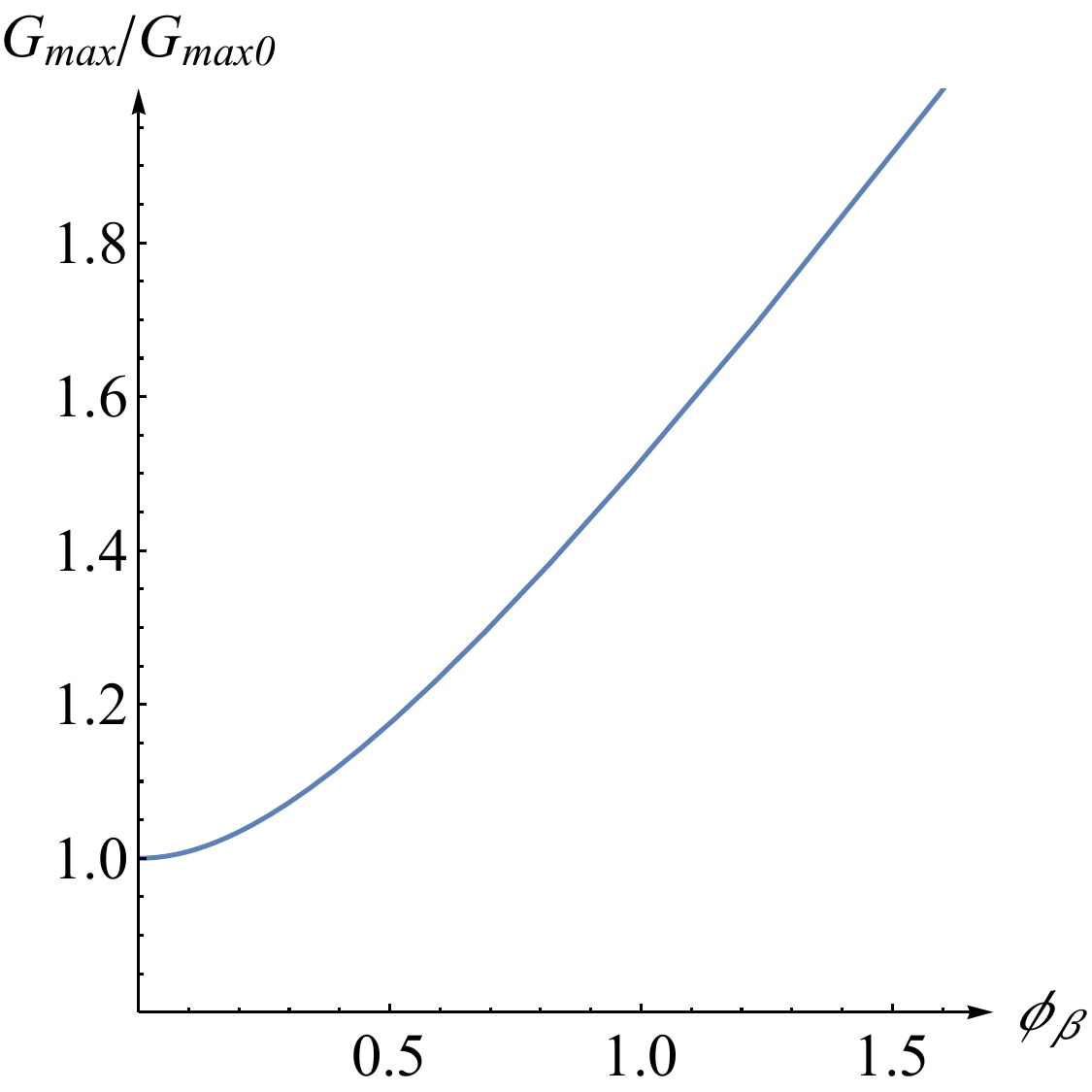}}\qquad
	\subfigure[$\mathcal{S}=0.25 $]{
		\includegraphics[width=0.4\textwidth]{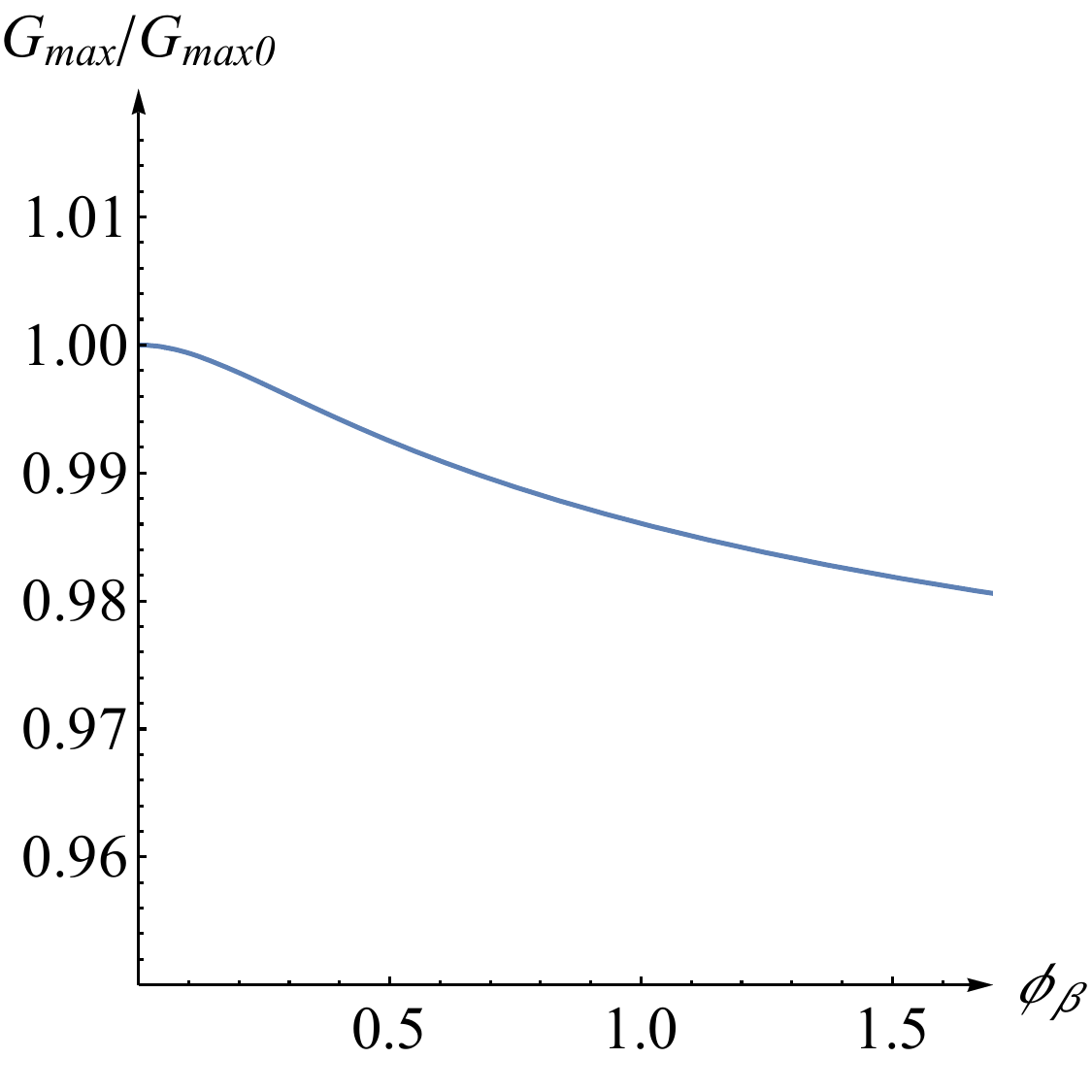}}\qquad
	\caption{For $\ell_{\rm AdS}=1$, relationship between growth rate and $\phi_{\beta}$. Fixed the $\mathcal{E}^{(\beta)}=1/16 \pi$,  $\mathcal{S}=0.25$ respectively, the larger the parameter $\phi_{\beta}$, the smaller the maximum value of the growth rate function $G(r)$. } \label{Fignull991}
\end{figure}

We perform similar numerical calculations as before, and the results are shown in the Figure \ref{Fignull991}. The Figure \ref{Fignull991}(a) shows that the growth rate will increase when the $\phi_{\beta}$ increase while keeping the $\mathcal{E}^{(\beta)}$ a constant. This means that the vacuum case has the minimum growth rate with fixed mass. The Figure \ref{Fignull991}(b) shows that the growth rate will decrease when the $\phi_{\beta}$ increase with a fixed entropy density $\mathcal{S}$. That is to say the vacuum case gives the maximum growth rate with fixed entropy density.

From these numerical results we obtain following two conclusions for the case when the dominant energy condition is broken and the bulk is not asymptotically Schwarzschild. Firstly, if we fix entropy density, the vacuum planar symmetric black hole will still have maximal entanglement growth rate. This indicates that the dominant energy condition and asymptotically Schwarzschild bulk geometry are not necessary conditions for our conclusion in the case of fixing entropy density. Further research needs to be done in the future to find the necessary and sufficient conditions. When we fix the energy density, the case become a little complicated. If we use standard quantization, we find the vacuum planar symmetric black hole will still have maximal entanglement growth rate. However, if we used alternative quantization, numerical results show that the matter will increase the maximal growth rate of entanglement. It is interesting to study if there is any deep physics resulting such difference in the future.

\section{Summary and Discussion}
In this paper, we have studied the upper bounds of the holographic entanglement entropy growth rate which calculated by the H-M surface.  The H-M surface is a very interesting bulk geometric structure. It can be used not only to calculate the entanglement entropy of the bipartition of the thermofield double state in this paper, but also to calculate the entanglement entropy of other choices of boundary subregions. In this paper, we focused on bipartition of the thermofield double state to simplify the discussion. The methods and teachinques used in this paper can be generalized into the entanglement entropy of other choices of boundary subregion straightforwardly. Similar to the codimension-2 H-M surface, the codimension-1 extremal surface can be used to calculate the holographic complexity of the thermofield double states. Its time evolution also has properties similar to H-M surfaces and an different upper bound has also been found in Refs.~\cite{Susskind:2014rva,Stanford:2014jda,Yang:2019alh}.

In order to obtain the upper bounds of growth rate, we first study the Schwarzschild-AdS black hole spacetime with planar, spherical and hyperbolic symmetries and find that there are three different maximums of the growth rate with given mass density, entropy density and temperature respectively. Then we study the RN-AdS black hole cases in the same way and find that the growth rates are always less than that in the Schwarzschild-AdS black hole cases with the same symmetries and given mass density, entropy density respectively. That is to say, the existence of Maxwell electromagnetic fields slows down the growth rate of holographic entanglement entropy if the black hole has spherical or planar symmetry.
	
The example of RN black hole inspires us to give a conjecture: For all static planar and spherically symmetric asymptotically Schwarzschild-AdS black holes of same mass density or entropy density, the vacuum AdS black hole gives the maximum entanglement entropy growth rate. We prove this conjecture by under dominant energy condition and Einstein' equation.  Furthermore, we considered the black hole spacetime with real scalar fields and take a negative ``mass-square'', which does not obey the dominant energy condition and the bulk geometry is not asymptotically Schwarzschild. We perform numerical calculations to give the relationships between the maximum growth rate and the parameters $\phi_{\alpha}$ and $\phi_{\beta}$. When we fix the total energy, under the standard quantization scheme, the vacuum black hole case will give the upper bound of the growth rate, and the maximum growth rate will decrease with the increase of parameter $\phi_{\alpha}$. However, when we choose alternative quantization scheme, the result will be opposite. The maximum growth rate will increase with the increase of parameter $\phi_{\beta}$, and the growth rate given by vacuum black hole is the smallest. When we fix the entropy density, in both quantization schemes, the vacuum black hole case gives the upper bound of the growth rate, and the maximum growth rate will decrease with the increase of parameter $\phi_{\alpha}$ or $\phi_{\beta}$.

We have gotten the conjecture that the vacuum black hole spacetime gives the upper bound of the  entanglement entropy growth rate with a given mass density or entropy. But with a given temperature, we have not discussed which case gives the upper bound. Our proofs in section \ref{section 4} have not discussed the cases with hyperbolic symmetry, but some examples we have shown in this paper suggest that the hyperbolic cases also follow the same conclusion. Our proofs are based on the dominant energy condition, but the result of the cases with real scalar field indicates that the dominant energy condition is not a necessary condition for our conclusion. For the cases that do not obey the dominant energy condition, we only consider the example of the AdS black hole with minimum coupled real scalar field to study the non vacuum case. We have not covered more complicated situations, there may be further discussed in our future work. Further research needs to be done in the future to find the necessary and sufficient conditions.

\acknowledgments
The work is supported by the NSFC(Natural Science Foundation of China) under Grant No.
12005155.

\appendix
\section{A proof of some properties of function $t_{B}(z_{A})$}\label{appendix A0}
In this appendix, we will prove: For $z_{A} = z_{\rm ext}$, $t_{B}\rightarrow \infty$, for $z_{A} \neq z_{\rm ext}$, $t_{B}$ is finite. This result is determined by the convergence of the integral \eqref{eq:40594} at the integral starting point $z_{A}$. We expand $G(z)$ around $z_{A}$ as
\begin{equation}
	G(z)=G(z_{A})+(z-z_{A})\frac{\partial G(z)}{\partial z}\bigg|_{z_{A}}+ \mathcal{O}((z-z_{A})^2).
\end{equation}
For the case of $z_{A}=z_{\rm ext}$, we have $\frac{\partial G(z)}{\partial z}\bigg|_{z_{\text{ext}}}=0$, so the expansion of $G(z)$ is
\begin{equation}
	G(z)=G(z_{\rm ext})+ \mathcal{O}((z-z_{\rm ext})^2).
\end{equation}
So when $z_{A} \neq z_{\rm ext}$, the integrand function of \eqref{eq:40594} can be expanded as
\begin{equation}
	\label{eq:40590001}
	\begin{aligned}
		&\frac{G(z_{A})}{f(z)\mathrm{e} ^{-\frac{\chi(z)}{2}}\sqrt{G(z_{A})^2+\frac{f(z)}{z^{2(d-1)}}}}
		\\
		=&\frac{G(z_{A})}{f(z)\mathrm{e} ^{-\frac{\chi(z)}{2}}\sqrt{G(z_{A})^2-G(z)^2}} \\
		=&\frac{G(z_{A})}{f(z)\mathrm{e} ^{-\frac{\chi(z)}{2}}\sqrt{2(z-z_{A})G_{1}+ \mathcal{O}((z-z_{A})^2)}}\\
		\sim & \frac{1}{\sqrt{(z-z_{A})}}
	\end{aligned}
\end{equation}
Where we have written $\frac{\partial G(z)}{\partial z}\bigg|_{z_{A}}=G_{1}$. Obviously, the integral of \eqref{eq:40594} is convergent around $z_{A}$ when $z_{A} \neq z_{\rm ext}$.

Similarly for $z_{A} = z_{\rm ext}$, the integrand function of \eqref{eq:40594} can be expanded as
\begin{equation}
	\label{eq:40590002}
		\frac{G(z_{A})}{f(z)\mathrm{e} ^{-\frac{\chi(z)}{2}}\sqrt{G(z_{A})^2+\frac{f(z)}{z^{2(d-1)}}}}
		=\frac{G(z_{A})}{f(z)\mathrm{e} ^{-\frac{\chi(z)}{2}}\sqrt{\mathcal{O}((z-z_{\rm ext})^2)}}
		\sim  \frac{1}{z-z_{\rm ext}}
\end{equation}
Obviously, the integral of \eqref{eq:40594} is divergent around $z_{A}$ when $z_{A} = z_{\rm ext}$.

\section{The area of extremal surface parameterized by continuous parameter}\label{appendix A}
\begin{figure}
	\centering
	\includegraphics[width=0.5\textwidth]{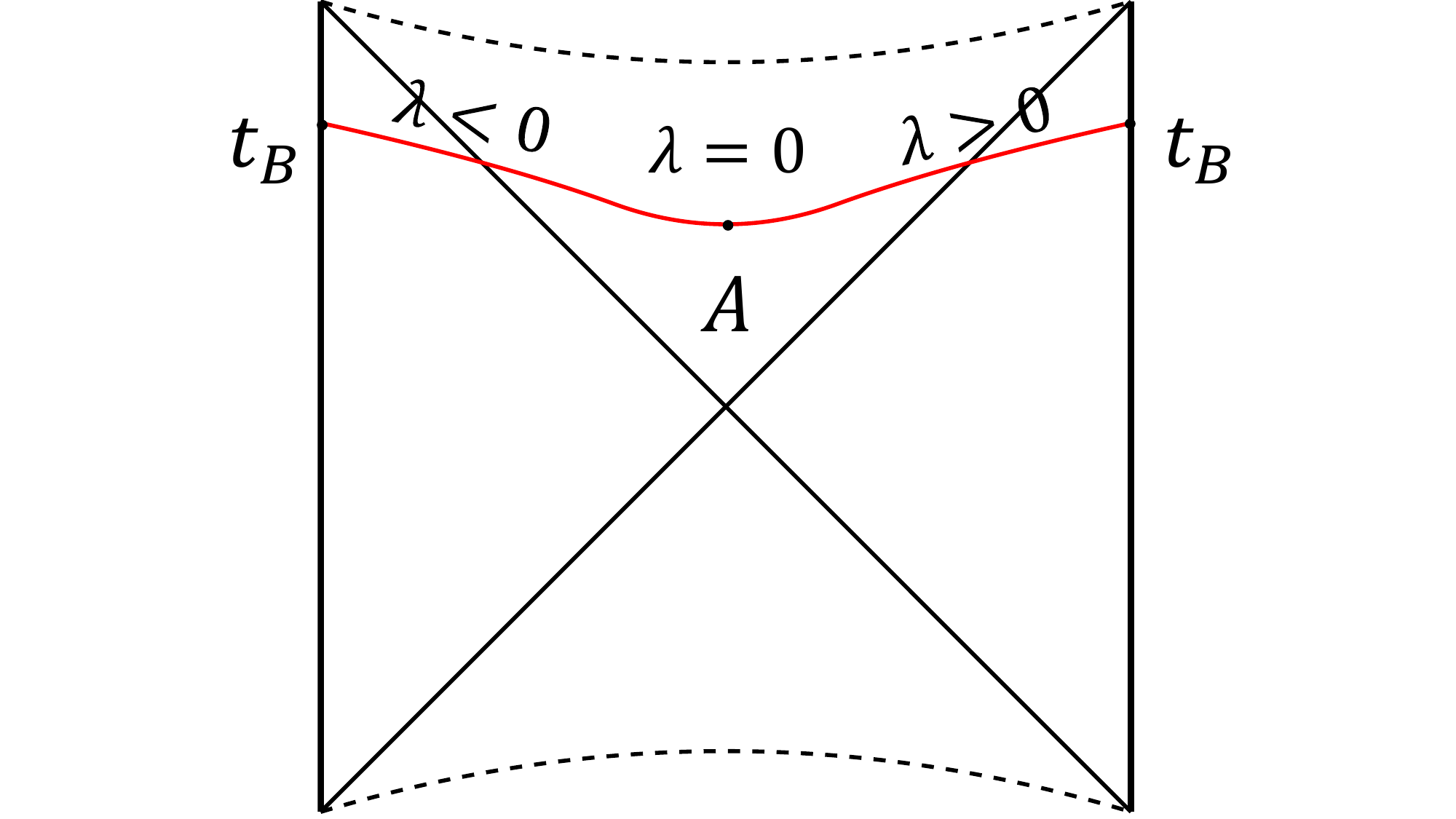}
	\caption{Set $t_{R}=t_{L}=t_{B}$. A is the middle point with $\lambda=0$.} \label{Fignulla13}
\end{figure}

We have calculated the area of extremal surface by using the $t$ coordinate as the integral variable in the section \ref{section 2}. The extremal surface is continuous at the horizon, but the $t$ coordinate is divergent. In this appendix, we will parameterize the extremal surface with a continuous real number $\lambda$, and get the same result as that in section \ref{section 2}.
The codimension-2 extremal surface $\Gamma$ can be parameterize by $t=t(\lambda,x^i)$ and $z=z(\lambda,x^i)$(see Figure \ref{Fignulla13}).  We set $\lambda \rightarrow -\infty$ and $\lambda\rightarrow +\infty$ at the left and right boundaries respectively, and $\lambda=0 $ at the "middle point" A(Where we have $t=0$.).
	
Since we can always let $t_{R}=t_{L}$ by Lorentz boost, the extremal surface will be symmetric in the left and right parts. At the middle point $A$($\lambda_{A}=0$), we will have $\partial z/\partial \lambda=0$ because of the symmetry and extremum constraints. The area $A(\Gamma)$ also can be computed by the right half part($0 \leqslant \lambda \leqslant +\infty $) multiplied by two. We use a dot to represent the derivative with respect to $\lambda$. The area of the extremal surface $\Gamma$ reads
\begin{equation}
		\label{eq:7054}
		\mathcal{A}(\Gamma)=2V_{k,d-2}\int_{0}^{+\infty} \td\lambda \frac{1}{z^{d-1}}\sqrt{-f\mathrm{e}^{-\chi}\dot{t}^2+f^{-1}\dot{z}^{2}}.
\end{equation}
Where $V_{k,d-2}=\int\td \sigma_{k,d-2}$ is the unit volume of the $d-2$ dimensional spatial directions, here $\td \sigma_{k,d-2}^{2}$ is the induced metric of  the constant of $x_{1}$ in $\td \Sigma_{k,d-1}^{2}$.
	
In order to get the growth rate of area, we can regard the area $\mathcal{A}$ as an action, and the integrand function of (\ref{eq:7054}) as a Lagrangian, which is a functional of generalized coordinate and generalized velocity $z,\dot{z},\dot{t}$
\begin{equation}
		\label{eq:70542}
		\mathcal{L} (z,\dot{z},\dot{t})=\frac{1}{z^{d-1}}\sqrt{-f\mathrm{e}^{-\chi}\dot{t}^2+f^{-1}\dot{z}^{2}}.
\end{equation}
According to classical mechanics, the partial derivative of action over generalized coordinate $\partial S/\partial q$ is generalized momentum, so the partial derivative of area over $t$(as a generalized coordinate) is $\partial \mathcal{A}/\partial t=2V_{k,d-2}\mathcal{P}$. Where $\mathcal{P}$ is the "generalized momentum" of the extremal surface.
At the boundary we have $z=0$, so that $t=t_{B}$, we get the rate of area growth over the boundary time
\begin{equation}
		\label{eq:70501}
		\frac{\td \mathcal{A}}{\td t_{B}}=2V_{k,d-2}\mathcal{P}|_{t=t_B,z=0}.
\end{equation}
Since the Lagrangian $\mathcal{L}$ does not depend explicitly on $t$, we can get a conserved quantity on the extremal surface
\begin{equation}
		\label{eq:7055}
		\mathcal{P}=\frac{\partial \mathcal{L}}{\partial \dot{t}}=\frac{-f\mathrm{e}^{-\chi}\dot{t}+f^{-1}\dot{z}}{z^{d-1}\sqrt{-f\mathrm{e}^{-\chi}\dot{t}^2+f^{-1}\dot{z}^{2}}}.
\end{equation}
Since the parameter $\lambda$ can be chosen freely, we take $\lambda$ as the length parameter, which satisfies $\mathcal{A}=2V_{k,d-2}\int \td\lambda $, thus
\begin{equation}
		\label{eq:7056}
		\frac{1}{z^{d-1}}\sqrt{-f\mathrm{e}^{-\chi}\dot{t}^2+f^{-1}\dot{z}^{2}}=1.
\end{equation}
Combining (\ref{eq:7055}) and (\ref{eq:7056}), eliminate $\dot{t}$, we can get
\begin{equation}
		\label{eq:7057}
		\mathcal{P}=\frac{-f\mathrm{e}^{-\chi}\sqrt{f^{-2}\mathrm{e}^{\chi}\dot{z}^2-f^{-1}\mathrm{e}^{\chi}z^{2(d-1)}}+f^{-1}\dot{z}}{z^{2(d-1)}}.
\end{equation}
In order to figure out the conserved quantity $\mathcal{P}$, we focue on the middle point A with $z=z_{A}$, there is an extra conditions $\dot{z}=0$, then we get
	
\begin{equation}
		\label{eq:7059}
		\mathcal{P}=\frac{\sqrt{-f(z_{A})\mathrm{e}^{-\chi(z_{A})}}}{z^{d-1}_{A}}.
\end{equation}
So we obtain the growth rate of the area
\begin{equation}
		\label{eq:70501}
		\frac{\td \mathcal{A}}{\td t_{B}}=2V_{k,d-2}\mathcal{P}=2V_{k,d-2}\frac{\sqrt{-f(z_{A})\mathrm{e}^{-\chi(z_{A})}}}{z^{d-1}_{A}}.
\end{equation}
The result is the same as (\ref{eq:405012}). The extremal surface is smooth at the horizon, where there is only coordinate singularity of $t$. However, it is easier for us to do the following calculations with the  $t$ coordinate as the integral variable in the section \ref{section 2}.

\bibliographystyle{JHEP}

\bibliography{ref-HEE-3}

\end{document}